%% file: axial.tex
\documentclass[12pt]{article}
\usepackage{latexsym,texdraw,epsf,a4}
\newcommand{\be}{\begin{equation}}
\newcommand{\ee}{\end{equation}}
\newcommand{\bea}{\begin{eqnarray}}
\newcommand{\eea}{\end{eqnarray}}
\newcommand{\deltafour}{\left( 2\pi \right)^4 \delta}
\newcommand{\gBBP}{g_{B^\ast B \pi}}
\newcommand{\mev}{\,\mathrm{Me\kern-0.1em V}}
\newcommand{\gev}{\,\mathrm{Ge\kern-0.1em V}}
\newcommand{\Tr}{\mathop{\mathrm{Tr}}}
\newcommand{\csw}{c_{\mathrm{sw}}}
\makeatletter

\let\errp\errparen
\long\def\@makecaption#1#2{%
  \vskip\abovecaptionskip
  \sbox\@tempboxa{\small{\bfseries #1} \  #2}%
  \ifdim \wd\@tempboxa >\hsize
    \small{\bfseries #1} \  #2\par
  \else
    \global \@minipagefalse
    \hb@xt@\hsize{\hfil\box\@tempboxa\hfil}%
  \fi
  \vskip\belowcaptionskip}
\makeatother
\begin{document}
\begin{titlepage}
\begin{center}
\huge\bfseries
Towards a lattice determination of the $B^\ast B \pi$ coupling
\end{center}
\vfill
\begin{center}
\large{\bfseries UKQCD Collaboration}\\[1ex]
G.M. de Divitiis$^{\mathrm a,}$\footnote{\tt giulia@hep.phys.soton.ac.uk}, 
L. Del Debbio$^{\mathrm a,}$\footnote{\tt ldd@hep.phys.soton.ac.uk},
M. Di Pierro$^{\mathrm a,}$\footnote{\tt mdp@hep.phys.soton.ac.uk},\\
J.M. Flynn$^{\mathrm a,}$\footnote{\tt j.flynn@hep.phys.soton.ac.uk}, 
C. Michael$^{\mathrm b,}$\footnote{\tt cmi@liv.ac.uk} and
J. Peisa$^{\mathrm b,}$\footnote{\tt J.J.Peisa@swansea.ac.uk}%
  $^,$\footnote{Present address: Dept.\ of Physics,
  Univ.\ of Wales Swansea, Singleton Park, Swansea SA2 8PP, UK}
  \\[2em]
$^{\mathrm a}$ Dept.\ of Physics and Astronomy, Univ.\ of Southampton,\\
Southampton SO17 1BJ, UK.\\[0.5em]
$^{\mathrm b}$ Theoretical Physics Division, Dept.\ of Mathematical
Science,\\ Univ.\ of Liverpool, Liverpool L69 3BX, UK.
\end{center}
\vfill
\begin{center}
\textbf{Abstract}
\end{center}
\begin{quote}
  The coupling $g_{B^\ast B \pi}$ is related to the form factor at
  zero momentum of the axial current between $B^\ast$- and
  $B$-states. This form factor is evaluated on the lattice using
  static heavy quarks and light quark propagators determined by a
  stochastic inversion of the fermionic bilinear. The $\gBBP$ coupling
  is related to the coupling $g$ between heavy mesons and low-momentum
  pions in the effective heavy meson chiral lagrangian. The coupling
  of the effective theory can therefore be computed by numerical
  simulations. We find the value $g = 0.42(4)(8)$. Besides its
  theoretical interest, the phenomenological implications of such a
  determination are discussed.
\end{quote}
\vfill
\begin{flushleft}
SHEP 98-09 \\
LTH-429 \\
20 July 1998
\end{flushleft}

\end{titlepage}
\setcounter{footnote}{0}

\section{Introduction}

The precise determination of the Cabibbo-Kobayashi-Mas\-ka\-wa (CKM)
matrix will provide a stringent consistency test of the Standard Model
(SM), together with new handles to understand CP violation and search
for hints of new physics. As far as the heavy flavour sector is
concerned, a large amount of experimental data is expected from
$B$-factories and CLEO in the near future. Non-perturbative QCD
effects are the main source of systematic error in the extraction of
fundamental parameters from experimental data: reliable results depend
on some way to tame the large-distance dynamics.

Lattice simulations provide a powerful tool to investigate
non-per\-tur\-ba\-ti\-ve dynamics from first principles, but
systematic errors are introduced by the finite lattice spacing and the
restricted range of quark masses which may be simulated. A different
approach is based on exploiting exact or approximate symmetries of the
theory to write effective lagrangians describing the large distance
behaviour in terms of effective meson fields. The chiral symmetry of
strong interactions in the limit where $m_q \rightarrow 0$ constrains
the terms appearing in the chiral lagrangian. The couplings in this
lagrangian are phenomenological inputs, which need to be taken either
from experimental data or from some other source. On the other hand,
when the quark masses tend towards infinity, heavy quark effective
theory (HQET) has proved to be a powerful tool to study heavy-flavour
physics. However, HQET is less powerful when applied to heavy-to-light
transitions, like $B \rightarrow \pi$, where the normalisation of
matrix elements is not fixed by the symmetry. A combination of these
two symmetries has been used in recent years to develop the heavy
meson chiral lagrangian (HM$\chi$), describing the interactions of
low-momentum pions with mesons containing a single heavy
quark~\cite{wise92} (for reviews see~\cite{grin-mex, casalbuoni97}).

The definitions and notations used throughout this paper are
conveniently introduced by a succint description of the building
blocks of this effective theory.

For the heavy degrees of freedom, heavy quark symmetry predicts that
the wave function of a heavy meson is independent of the flavour and
spin of the heavy quark, leading to a covariant representation of
heavy mesonic states~\cite{neubert}.  The angular momentum $j$ and the
parity $P$ of the light degrees of freedom determine a degenerate
doublet of states with spin--parity $J^P = (j \pm 1/2)^P$. The
pseudoscalar and vector mesons (e.g.\ $B$ and $B^\ast$) correspond to
the $j=1/2$ case and are described by a $4\times 4$ Dirac matrix $H$
with two spinor indices, one for the heavy quark and one for the light
degrees of freedom. In terms of the effective meson fields:
\begin{equation}
        H = \frac{1 + v \!\!\!/}{2}\, 
        \left[B^\ast_\mu \gamma^\mu - B \gamma_5\right],\qquad
        \overline H = \gamma^0 H^\dagger \gamma^0
\end{equation}
where $v$ is the velocity of the meson and $B$ and $B^\ast$ are the
annihilation operators for particles containing a $b$ quark in the
initial state.  These meson fields are used in the effective
lagrangian to describe the heavy mesons. The light mesons are treated
as an octet of pseudo-Goldstone bosons according to the usual chiral
lagrangian formalism. At low-momentum the strong interactions of $B$
and $B^\ast$ mesons with light pseudoscalars are described by the
couplings in the effective lagrangian; the lowest order interaction is
given by~\cite{wise92,casalbuoni97}:
\begin{equation}
\label{eq:def_g}
        {\mathcal{L}}_{\mathrm{HM}\chi}^{\mathrm{int}}= g 
        \Tr
        (\overline{H}_a H_b {\mathcal{A}}^{ba}_\mu \gamma^\mu \gamma^5)
\end{equation}
where 
\begin{equation}
        {\mathcal{A}}_\mu =
        {i\over2}
           (\xi^\dagger\partial^\mu \xi - \xi\partial^\mu \xi^\dagger)
\label{eq:amu}
\end{equation}
with $\xi=\exp(i{\mathcal{M}}/f)$. $\mathcal{M}$ is a $3\times 3$
matrix of $\pi$, $\eta$ and $K$ fields
\begin{equation}
\mathcal{M} = \left(\begin{array}{ccc}
\pi^0/\sqrt2 + \eta/\sqrt6 & \pi^+                       & K^+ \\
\pi^-                      & -\pi^0/\sqrt2 + \eta/\sqrt6 & K^0 \\
K^-                        & \bar K^0           & -\sqrt{2/3}\eta
		    \end{array}\right)
\end{equation}
and the trace is over Dirac indices. At tree level $f$ can be set
equal to $f_\pi$ (the definition of the decay constant used here sets
$f_\pi = 132\mev$). The roman indices $a$ and $b$ denote light quark
flavour and repeated indices are summed over $1,2,3$. The expansion of
${\mathcal A}$ in terms of pion fields begins with a linear term,
\begin{equation}
        {\mathcal{A}}_\mu = - \frac1{f}\partial_\mu
        {\mathcal{M}} +
        \cdots
\end{equation}

The coupling $g$ in Eq.~\ref{eq:def_g} can easily be related to the
$B^\ast B \pi$ coupling defined as~\cite{yan92,BBKR}:
\begin{equation}
\label{eq:def_G}
        \langle B^0(p) \pi^+(q) | B^{\ast +}(p^\prime) \rangle = 
        - g_{B^\ast B \pi}(q^2) q_\mu \eta^\mu(p')
        \left( 2\pi \right)^4 \delta(p^\prime - p -q)
\end{equation}
where $\eta^\mu$ is the polarisation vector of the $B^\ast$ and
the physical states are relativistically normalised:
\begin{equation}
        \langle B(p) | B(p^\prime) \rangle =
         2 p^0 (2\pi)^3 \delta^{(3)}(\bf p - \bf p^\prime)
\end{equation}
The physical coupling $g_{B^\ast B \pi}$ is given by the value of the
above form factor for an on-shell pion:
\begin{equation}
        g_{B^\ast B \pi} = \lim_{q^2 \rightarrow m_\pi^2}
        g_{B^\ast B \pi}(q^2)
\end{equation}
At tree level in the heavy meson chiral lagrangian, the
above matrix element is:
\begin{equation}
        \langle B^0(p) \pi^+(q) | B^{\ast +}(p^\prime) \rangle = 
        - \frac{2 m_B}{f_\pi}\, g \, q_\mu \eta^\mu(p^\prime)
        \deltafour(p^\prime - p - q)    
\end{equation}
which therefore yields:
\begin{equation}
\label{eq:g_to_G}
        g_{B^\ast B \pi} = \frac{2 m_B}{f_\pi} g
\end{equation}
As a result, $g$ and $g_{B^\ast B \pi}$ are considered as equivalent
throughout this paper. The above relation can be extended to take into
account higher-order terms in the HM$\chi$
lagrangian~\cite{ccllyy,fleischer93,falkgrin94}.

Starting from Eq.~\ref{eq:def_G} and performing an LSZ reduction of
the pion field, the coupling $g$ is related to the form factor at zero
momentum-transfer of the axial current between hadronic states. Such a
relation is the analog, in the $B \pi$ system, of the
Goldberger-Treiman relation, relating the nucleon electromagnetic form
factor to the nucleon-nucleon-pion coupling. An important consequence
of the Goldberger-Treiman relation, for our purposes, is that it
allows a numerical evaluation of the $g_{B^\ast B \pi}$ coupling, as
the form factors of the axial current can be evaluated by a lattice
simulation. The details of the pion reduction are presented in
Section~\ref{sec:lsz}.

The interest of such a computation is two-fold. From a theoretical
point-of-view, it is interesting {\it per se} to be able to fix, from
lattice QCD, the coupling appearing in the heavy meson chiral
lagrangian. On the other hand, it is important to stress that this
determination also has phenomenological motivations. Assuming vector
dominance in the $B\rightarrow \pi l \nu$ decay, the coupling
$g_{B^\ast B\pi}$ fixes the normalisation of the form factors used to
parametrise the matrix element of the weak vector current, $V^\mu =
\bar u \gamma^\mu b$, between hadronic states. Defining the form 
factors by:
\begin{eqnarray}
        \langle \pi^+(p^\prime) | V^\mu | \bar B(p) \rangle &=&
        f_1(q^2) (p+p^\prime - \frac{m_B^2 - m_\pi^2}{q^2} q)^\mu +
        f_0(q^2) \frac{m_B^2 - m_\pi^2}{q^2} q^\mu \nonumber \\ &=&
        f_+(q^2) (p+p^\prime)^\mu + f_-(q^2) q^\mu
\end{eqnarray}
where $q=p-p^\prime$ is the transferred momentum, the contribution
from the $B^\ast$ channel is easily evaluated:
\begin{equation}
        f_1(q^2) = \frac{g_{B^\ast B \pi}}{2 f_{B^\ast}}
        \frac{1}{1 - q^2/m_{B^\ast}^2}
\label{eq:pole}
\end{equation}
The normalisation of the form factor therefore depends on the $B^\ast
B \pi$ coupling and the decay constant of the vector meson, defined
as:
\begin{equation}
        \langle 0 | V^\mu | \bar B^\ast(p) \rangle = 
        i
        \frac{m^2_{B^\ast}}{f_{B^\ast}} \eta^\mu(p)
\end{equation}
Heavy quark symmetry and chiral symmetry justify this pole form for
$f_1$ when $q^2$ is close to $q^2_{\mathrm{max}} = (m_B -
m_\pi)^2$~\cite{wise92,%
ccllyy,fleischer93,falkgrin94,boydgrin95,Burdman94}.  For $q^2$ far
from $q^2_{\mathrm{max}}$, the pole form may be taken as a
phenomenological ansatz.  However, we note that the functional
dependence of the form factor in Eq.~\ref{eq:pole} cannot be
simultaneously consistent with heavy quark symmetry at large $q^2$,
which demands that $f_1(q^2_{\mathrm{max}}) \sim m_B^{1/2}$ and the
light-cone sum rule scaling relation at $q^2=0$, which states
$f_1(q^2{=}0) \sim m_B^{-3/2}$~\cite{braun:rostock}.  Nonetheless, by
fitting lattice results, which are available in the high $q^2$ region
where the pole form is justified, we can determine the parameters in
Eq.~\ref{eq:pole}.

It is interesting to remark that the interplay of the effective
lagrangian approach and lattice simulations provides another
determination of the form factors for the heavy-to-light $B$ decays
and therefore sheds further light on the theoretical determination of
the non-perturbative effects mentioned at the beginning. This result
can be compared with direct computations of the same quantities
obtained by fitting lattice data~\cite{ldd97}, using unitarity
bounds~\cite{ll96} and sum rules~\cite{BBKR,sum}.

In the work described here, the matrix element of the light quark
axial current between the heavy mesons is computed in a quenched
lattice simulation of QCD in the static heavy quark limit, using
stochastic methods to compute the desired light quark propagators, as
described in Section~\ref{sec:lattice}. The discussion of systematic
errors is an important issue in any lattice calculation and plays a
crucial part in estimating the error on the final result. In this
respect, it is important to stress here that we are presenting an
exploratory study. Our main concern is therefore to test the
possibility of extracting the coupling defined above, rather than
presenting its best lattice determination. Such a task would require a
more extensive simulation and is left for further studies.

Renormalisation constants are needed in order to connect lattice
results with continuum physical observables. Those relevant for the
action and the quantities considered in this paper are summarised in
Sect.~\ref{sec:ren}.

The best estimate we obtain for $g$ is
\begin{equation}
        g=0.42(4)(8)
\end{equation}
The phenomenological implications of this result are discussed in
Sec.~\ref{sec:pheno}.

\section{Pion reduction}
\label{sec:lsz}

An LSZ reduction of the pion in the definition of $g_{B^\ast B \pi}$,
Eq.~\ref{eq:def_G}, yields:
\begin{equation}
\label{eq:lsz1}
        \langle B^0(p) \pi^+(q) | B^{\ast +}(p+q) \rangle = 
        i (m_\pi^2 - q^2) \int_x e^{i q \cdot x}  
        \langle \bar B(p) | \pi(x) | B^\ast(p+q) \rangle 
\end{equation}
Using the PCAC relation:
\begin{equation}
        \pi(x) = \frac{1}{m_\pi^2 f_\pi} \partial^\mu A_\mu(x)
\end{equation}
where $A_\mu$ is, as usual, the QCD axial current, Eq.~\ref{eq:lsz1}
becomes:
\begin{equation}
        \langle B^0(p) \pi^+(q) | B^{\ast +}(p+q) \rangle = 
        q^\mu \frac{1}{f_\pi} \frac{m_\pi^2 - q^2}{m_\pi^2}
        \int_x e^{i q\cdot x}  
        \langle \bar B(p) | A_\mu(x) | B^\ast(p+q) \rangle 
\end{equation}
The matrix element of the axial current is parametrised in terms of
three form factors:
\begin{eqnarray}
        \langle B^0(p) | A_\mu(0) | B^{\ast +}(p+q) \rangle &=&
        \eta_\mu F_1(q^2) +
        \left(\eta \cdot q \right) (2p + q)_\mu F_2(q^2) 
        \nonumber \\
        &&+ \left(\eta \cdot q \right) q_\mu F_3(q^2)
\label{eq:axial_ff}
\end{eqnarray}
yielding for the $B^\ast B \pi$ coupling:
\begin{equation} 
\label{eq:tbac}
        g_{B^\ast B \pi}(q^2) = -
        \frac{1}{f_\pi} \frac{m_\pi^2 - q^2}{m_\pi^2}
        \left[F_1(q^2) + \left(m_{B^\ast}^2 - m_B^2\right) F_2(q^2)
        + q^2 F_3(q^2)\right]
\end{equation}
In the static limit in which our simulation is performed, the $B$ and
$B^\ast$ mesons are degenerate, so that the form factor $F_2$ can be
discarded.

Analytical continuation of Eq.~\ref{eq:tbac} towards the soft--pion
limit ($q^2 \rightarrow 0$), leads to:
\begin{equation}
\label{eq:GT}
        g_{B^\ast B \pi}(0) = - \frac{1}{f_\pi} F_1(0)
\end{equation}
It is commonly assumed, when deriving the Goldberger-Treiman relation,
that $g_{B^\ast B \pi}$ is a smooth function of $q^2$, and, therefore,
that the physical coupling can be approximated by:
\begin{equation}
        g_{B^\ast B \pi} = g_{B^\ast B \pi}(m_\pi^2) 
        \approx g_{B^\ast B \pi}(0) 
\end{equation}
The above equation explicitly shows that, in the soft--pion limit, the
$B^\ast B \pi$ coupling is related to the form factor of the axial
current between $B$ and $B^\ast$ states. If one were working in the
chiral limit from the very beginning, the same Goldberger-Treiman
relation, Eq.~\ref{eq:GT}, would be obtained from the conservation of
the axial current.

The relation between $g_{B^\ast B \pi}$ and $g$ mentioned earlier can
be rederived by comparing the matrix element of the Noether current
associated to chiral symmetry both in HM$\chi$ and QCD. In the chiral
limit, the Noether currents associated with chiral symmetry, in QCD
and in HM$\chi$, are respectively
\begin{equation}
        j^{5\;ab}_{{\mathrm{QCD}}\;\mu} 
        = \bar q^a\gamma_\mu \gamma_5 q^b
\end{equation}
and
\begin{equation}
        j^{5\,ab}_{{\mathrm{HM}}\chi\;\mu}
         = f_\pi \partial_\mu {\mathcal{M}}^{ab} 
        - 2 g
        \left( B^{\ast\,a\;\dagger}_\mu B^b +
               B^{a\;\dagger} B^{\ast~b}_\mu 
        \right)
        + \cdots
\label{eq:eff_axial}
\end{equation}
where the ellipsis denotes terms with more than one pion or terms
containing both heavy-mesons and pions.

\begin{figure}
\centerline{\setlength\epsfxsize{0.9\hsize}
\epsfbox{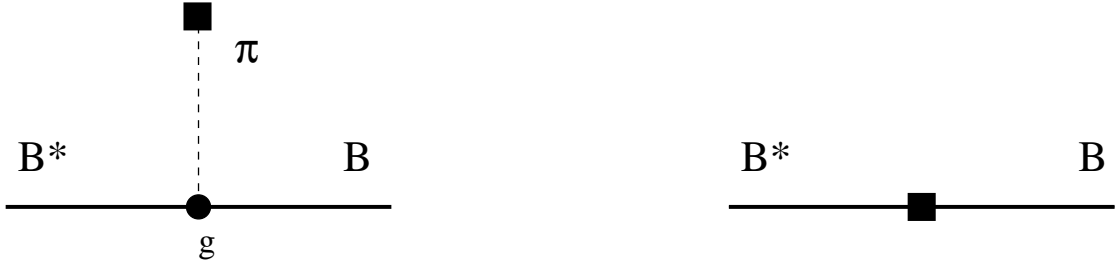}}
\caption{Tree-level diagrams needed to compute the matrix element of
the axial current in HM$\chi$. The dot ($\bullet$) represents the
$g$-vertex in the HM$\chi$ lagrangian; the squares are insertions of
the axial current in Eq.~\ref{eq:eff_axial}.}
\label{fig:tree_diag}
\end{figure}

The form factors of the axial current in QCD are related to the
coupling of the heavy meson chiral lagrangian by matching the two
theories at tree level. The diagrams needed at tree level to evaluate
the matrix element in Eq.~\ref{eq:axial_ff} for the HM$\chi$ current
are depicted in Fig.~\ref{fig:tree_diag}. A straightforward
computation leads to:
\begin{eqnarray}
        F_1(q^2) &=& - 2 g \, m_B \nonumber \\
        F_3(q^2) &=& \frac{2g}{q^2} m_B \nonumber
\end{eqnarray}
which, together with the Golberger-Treiman relation, reproduces
Eq.~\ref{eq:g_to_G}. One may also work away from the chiral and heavy
quark limits and include corrections for finite mass pions and heavy
quarks in this result.

The determination of $g$ is therefore reduced to the computation of
the matrix element of the light-light axial vector current between
hadronic states. Such an evaluation can be performed using three-point
correlation functions on the lattice. The details of the calculation
are reported in the next section.

\section{Lattice results}
\label{sec:lattice}

\subsection{Stochastic propagators}

The numerical analysis is carried out on 20 quenched gauge
configurations, generated on a $12^3 \times 24$ lattice at
$\beta=5.7$, corresponding to $a^{-1}=1.10$ GeV. The heavy quark
propagators are evaluated in the static limit~\cite{eichten88-89}.
Stochastic propagators~\cite{michael98,roma-tv96} are used to invert
the fermionic matrix for the light quarks. They can be used in place
of light quark propagators calculated with the usual deterministic
algorithm.  The stochastic inversion is based on the relation:
\begin{equation}
        S_{ij} =  M_{ij}^{-1}=\frac1Z \int {\cal D}\phi\;
        (M_{jk}\phi_k)^\ast \phi_i\; 
        \exp \left( -\phi_i^\ast (M^\dagger M)_{ij}
        \phi_j \right) 
\end{equation} 
where, in our case, $M$ is the tadpole improved SW fermionic operator
and the indices $i,j,k$ represent simultaneously the space-time
coordinates, the spinor and colour indices. Two values of $\kappa$ are
considered, $\kappa_1=0.13843$ and $\kappa_2=0.14077$, with
$\csw=1.57$. The lighter value $\kappa_2$ corresponds to a bare mass
of the light quark around the strange mass. The chiral limit
corresponds to $\kappa_c=0.14351$~\cite{hugh97}.  For every gauge
configuration, an ensemble of 24 independent fields $\phi_i$ is
generated with gaussian probability:
\begin{equation} 
P[\phi] =\frac1Z \exp \left(
-\phi_i^\ast (M^\dagger M)_{ij} \phi_j \right) 
\end{equation} 
All light propagators are computed as averages over the
pseudo-fermionic samples:
\begin{equation}
        S_{ij} = 
        \left\{\begin{array}{l}
               \langle (M\phi)_j^\ast \phi_i \rangle \\
               \mathrm{or} \\
               \langle \left(\gamma_5 \phi^\ast\right)_j 
	               \left(M\phi\gamma_5\right)_i \rangle 
               \end{array}
        \right.
\label{eq:stock}
\end{equation}
where the two expressions are related by $S = \gamma_5 S^\dagger
\gamma_5$.  Moreover, the maximal variance reduction method is applied
in order to minimise the statistical noise~\cite{michael98}. Maximal
variance reduction involves dividing the lattice into two boxes
($0<t<T/2$ and $T/2<t<T$) and solving the equation of motion
numerically within each box, keeping the spinor field $\phi$ on the
boundary fixed.  According to the maximal reduction method, the fields
which enter the correlation functions must be either the original
fields $\phi$ or solutions of the equation of motion in disconnected
regions.  The stochastic propagator is therefore defined from each
point in one box to every point in the other box or on the
boundary. For this reason, when computing a three-point correlation
function
\begin{equation} 
\langle 0 | J(t_1,x) {\cal O}(t_0, y) J^\dagger(t_2,z) | 0 \rangle
\end{equation} 
one operator --- ${\mathcal O}$ in the present work --- is forced to
be on the boundary ($t_0=0$ or $T/2$) and the other two must be in
different boxes, while the spatial coordinates are not constrained.
If $j$ is a point of the boundary, not all the terms in $(M\phi)_j$
lie on the boundary because the operator $M$ involves first neighbours
in all directions. Hence, whenever a propagator $S_{ij}$ is needed
with one of the points on the boundary, we use whichever of the two
expressions in Eq.~\ref{eq:stock} has the spinor $M\phi$ computed away
from the boundary.

In smearing the hadronic interpolating operators, spatial fuzzed links
are used. Following the prescription in~\cite{michael98, michael95},
to which the interested reader should refer for details, the fuzzed
links are defined iteratively as:
\begin{equation}
        U_{\mathrm{new}} = {\mathcal P} \left(f U_{\mathrm{old}} 
          + \sum_{i=1}^4 U_{\mathrm{bend},i} \right)
\end{equation}
where ${\mathcal P}$ is a projector over $SU(3)$ and
$U_{\mathrm{bend},i}$ are the staples attached to the link in the
spatial directions. We take $f=2.5$ and use two iterations with fuzzed
links of length one. The smeared fermionic fields are defined
following~\cite{michael95}.

\subsection{Lattice computation}

In order to extract $g$, the three-point correlation function $C_3$
and the two-point correlation function $C_2$ for local ($L$) and
fuzzed ($F$) sources are computed. The $FF$ three-point function is
defined as:
\begin{equation}
        C^{FF}_{3~\mu\nu}({\bf x}; t_1,t_2) = 
        \frac1V \int d^3y\;
        \langle 0 | J_\nu^{B^\ast}({\bf y},-t_1)
        A^\mu({\bf x+y},0) J^{B~\dagger}({\bf y},t_2) |0\rangle
\end{equation}
where $J_\nu^{B^\ast}$ and $J^B$ are fuzzed operators with the quantum
numbers corresponding respectively to the $B^\ast$ and $B$
states. Analogous definitions hold for the $FL$ and $LL$ cases. For
time separations that are large enough to isolate the lowest energy
states, the three-point function is related to the axial current
matrix element:
\begin{eqnarray}
        C^{FF}_{3~\mu\nu}({\bf x}; t_1,t_2) 
        &\rightarrow&
        \langle 0 |  J_\nu^{B^\ast} | B^\ast_r \rangle
        \frac{\langle B^{\ast}_r | A^\mu({\bf x}) |B\rangle}{2m_B \, 2m_B} 
        \langle B |  J^{B~\dagger} | 0 \rangle
        e^{-M_B (t_1+t_2)} \nonumber \\
        &=& 
        Z^F \eta_\nu^r(0)
        \frac{\langle B^{\ast}_r | A^\mu({\bf x}) |B\rangle}{2m_B} 
        Z^F
        e^{-M_B (t_1+t_2)} 
        \label{eq:3pt} 
\end{eqnarray}
where $r$ is the polarisation label of the vector particle, and the
sum over polarisations is omitted.  In the static limit considered in
this paper, the slope of the exponential time-decay is not the
physical mass of the mesons; it can be interpreted as a binding
energy. Moreover, the two-point functions for the vector and
pseudoscalar particles are degenerate, leading to the same binding
energies and $Z$ factors for both the $B$ and $B^\ast$.  In order to
avoid confusion, the physical mass is denoted by $m_B$ and the binding
energy by $M_B$. $Z^F$ is the overlap of the interpolating operator
with the physical state, defined from the two-point functions:
\begin{eqnarray}
        C_2^{FF}(t) 
        &=& 
        \frac1V \int_V d^3y\; 
        \langle 0|J^B({\bf y},0) J^{B~\dagger}({\bf y},t) |0\rangle 
        \nonumber \\
        &\rightarrow& (Z^F)^2 e^{-M_B t}
\end{eqnarray}

Integrating the three-point function over $\bf x$ gives the matrix
element for zero momentum transfer. In this limit, the latter can be
expressed in terms of the form factor $F_1(q^2)$ in
Eq.~\ref{eq:axial_ff}.

The sum over polarisations in Eq.~\ref{eq:3pt} yields:
\begin{equation}
        \int d^3x\; C^{FF}_{3~\mu\nu}({\bf x}; t_1,t_2) 
        \rightarrow
        (Z^F)^2 (g_{\mu\nu} - \frac{p_\mu p_\nu}{m_B^2})
        F_1(0)
        e^{-M_B (t_1+t_2)} 
\end{equation}
The last equation shows that the three-point functions with $\mu
\neq \nu$ and $\mu = \nu = 0$ should vanish. Therefore, only the
correlators with $\mu=\nu=1,2,3$ are henceforth considered.

Moreover, taking rotational symmetry into account, $C^{FF}_3({\bf x};
t_1,t_2)$ is expected to be a function of the distance $r$ only, up to
cut-off effects. The three-point functions measured on the
lattice are averaged over equivalent $\bf x$ positions\footnote{The
symbol $$ \overline{\sum}_r^x $$ indicates the average on all the
spatial positions on the lattice compatible with the constraint
$|x|=r$. On a finite lattice $V=L^3$, only some distances are allowed
$$r=\sqrt{x_1^2+x_2^2+x_3^2}$$ because each $x_i$ must be an integer
between 0 and L/2.  For each allowed distance $r$, a given number of
terms $N_r$ appears in the above sum, yielding a relative error on
each point $\delta E_\mu(r,t) \sim N_r^{-1/2}$, e.g.
\[
N_0=1;~N_1=6;~N_{\sqrt{2}}=12;~N_{\sqrt{3}}=6
;~\ldots;~N_{\sqrt{108}}=1 
\]
}. The desired matrix element is obtained from the ratio:
\begin{equation}
        E_{\mu}(r,t) \stackrel{\mathrm{def}}=
        (Z^F)^2 \overline{\sum}_r^x 
        \frac{C^{FF}_{3~\mu\mu}({\bf x},t,t)}{C_2^{FF}(t) C_2^{FF}(t)}
        \rightarrow
        \frac{\langle B^{\ast} | A_\mu(r) |B \rangle}
        {2m_B}\, \eta_\mu 
\end{equation}
where the time-dependence cancels when the three-point function is
divided by the product of two-point functions.
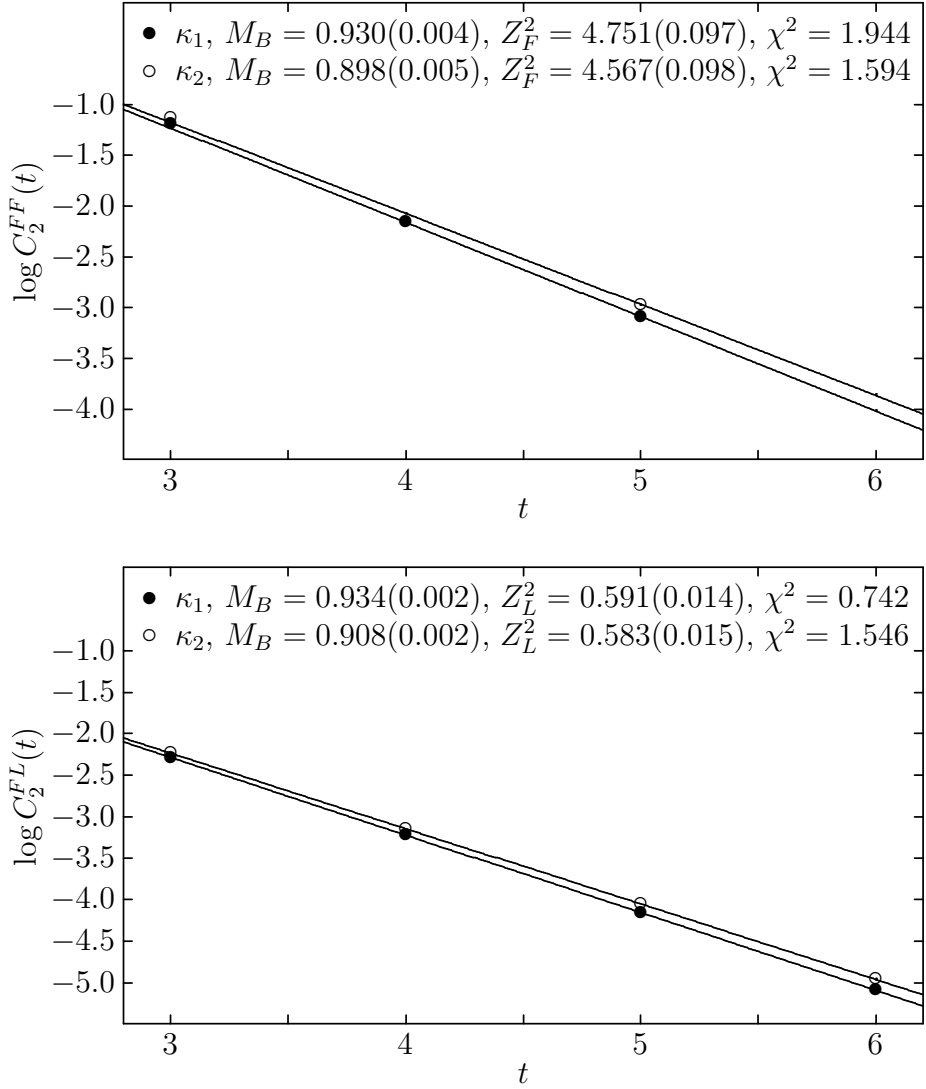
\begin{figure}
\input{fig2.tex}
\caption{Logarithmic plots of $C_2^{FF}(t)$ and $C_2^{FL}(t)$ for both
values of $\kappa$. The quoted values refer to the reduced $\chi^2$. 
\label{fig:c2}}
\end{figure}

The coefficients $Z^F$ and $Z^L$ are extracted from the exponential
fit of the two point correlation functions $C_2$. The data are
reported in Fig.~\ref{fig:c2}. As one can see from the plots, the
two-point functions already exhibit a single-exponential behaviour at
moderate time separations.  The main sources of error in this
computation are expected to stem from the determination of the
three-point function and from the value of the light quark masses,
which are far from the chiral limit. Thus, a single exponential fit of
the two-point functions turns out to be accurate enough for the scope
of this study.  The value of $Z^F$ is extracted from a direct fit of
$C^{FF}_2$, while $Z^L$ is obtained from $C_2^{FF}$ and
$C_2^{FL}$. The results of the fit are shown directly on the plot. It
is worth remarking that, in all the channels considered,
single-exponential fits yield reasonable values for the reduced
$\chi^2$.

The $B$ meson decay constant, in the static approximation, can be
extracted from:
\begin{equation}
        f_B^{\mathrm{static}} 
          = Z_A^{\mathrm{static}} \sqrt{\frac{2}{m_B}} Z^L a^{-3/2}
\end{equation}
where $Z_A^{\mathrm{static}}$ is the renormalisation constant for the
axial current in the static theory, which is discussed in the
following section and defined in Eq.~\ref{eq:ZAstatic}.  The aim here
is not a precise determination of the pseudoscalar decay constant,
$f_B^{\mathrm{static}}$. Rather, the result is presented to allow an
estimate of the systematic errors in our computation of the HM$\chi$
coupling, $g$.

The results of the fits, together with the values for
$f_B^{\mathrm{static}}$ are summarised in Tab.~\ref{tab:fb}.  The $B$
meson binding energies obtained from the fits of $C_2^{FF}$ and
$C_2^{FL}$ are consistent with each other. However, our determination
appears to be slightly different from the one published
in~\cite{michael98}.  The discrepancy can be explained as a
contribution from excited states, which is subtracted
in~\cite{michael98} where a multi-exponential fit is performed. The
1-state fit yields a value of $Z^L$ approximately 20\% higher than the
one obtained from the 3-state fit. Hence the value obtained for the
static $B$ decay constant lies above other lattice calculations of
this quantity~\cite{wittig97}. It is striking that the actual number
does not depend on the value of the hopping parameter $\kappa$.
However the variation, as the bare mass of the quark goes from $m_s$
towards the chiral limit, is also expected to be about 20\% and could
be obscured by the contamination from excited states.
\begin{table}
\begin{center}
\begin{tabular*}{\hsize}{@{}@{\extracolsep{\fill}}llll} 
\hline
                             & $\kappa_1$ & $\kappa_2$ & $\kappa_c$ \\[0.3ex]
\hline
$M_B a$                         & 0.930(4)   & 0.898(5)   & 0.862(7)   \\
$(Z^F)^2$                       & 4.75(10)   & 4.57(10)   & 4.37(15)   \\
$(Z^L)^2$                       & 0.59(1)    & 0.58(2)    & 0.57(3)    \\
$f_B^{\mathrm{static}}(\!\gev)$ & 0.43(1)    & 0.43(1)    & 0.42(2)    \\
\hline
\end{tabular*}
\end{center}
\caption{Values for $Z^F$, $Z^L$, $M_B$ and $f_B^{\mathrm{static}}$
obtained from fitting the two-point functions.
\label{tab:fb}}
\end{table}

This is a first, exploratory, direct lattice determination of the
coupling $g$, so the discrepancies noted above are perhaps expected
and could easily arise from various lattice artefacts. Our value of
$\beta$ is far from the continuum limit, while our action and
operators are not fully $O(a)$ improved. We have not tried to optimise
the smearing or fitting procedures. Our ensemble of gauge
configurations and the collection of spinor configurations on each
gauge sample are quite small. The calculation is also performed in the
quenched approximation. All of these issues could be addressed in a
further simulation, but here we will keep in mind that the
uncertainties will propagate as systematic errors to our final result
for the $B^*B\pi$ coupling.
  
The generation of stochastic propagators for the light quark and the
static approximation for the heavy quark have proved to be useful
tools in this investigation. However, neither is strictly necessary to
calculate the axial current matrix element of interest. One could
combine static heavy quarks with light quark propagators determined by
standard deterministic methods.  A full $O(a)$-improved simulation
with heavy quark masses around the charm mass would allow one to go
beyond the static approximation and study the dependence of $g$ on the
heavy mass.  The freedom allowed in lattice calculations to tune quark
masses would also allow the light quark mass dependence, noted above
as strikingly absent, to be investigated in more detail.

\begin{figure}
\input{fig3.tex}
\caption{Plots of $E^{0}(r,t)$ and $\overline{E}(r,t)$ as functions of
$r$ for different values of $t$ ($=3,4,5,6$) for $\kappa=0.13843$. 
\label{figc3a}}
\end{figure}
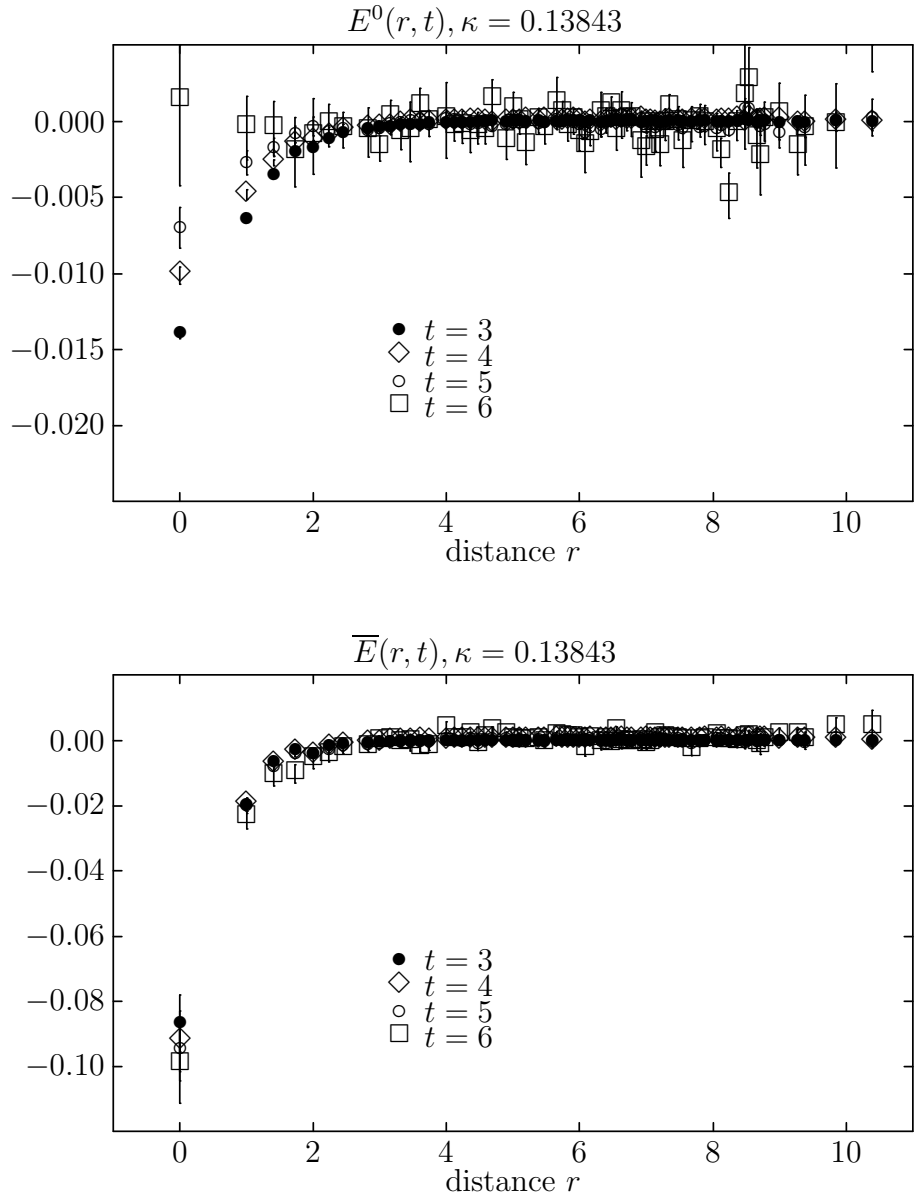

\begin{figure}
\input{fig4.tex}
\caption{Plots of $E^{0}(r,t)$ and $\overline{E}(r,t)$ as functions of
$r$ for different values of $t$ ($=3,4,5,6$) for $\kappa=0.14077$. 
\label{figc3b}}
\end{figure}

Returning to the results of the present simulation, the quantity
$E_{0}(x,t)$, which is expected to vanish, is measured as a further
consistency check.  The data reported in Figs.~\ref{figc3a}
and~\ref{figc3b} show a much smaller signal than the one obtained for
${E}_{i}(x,t)$. As $t$ is increased, the fictitious peak at zero
distance decreases, while the noise increases.

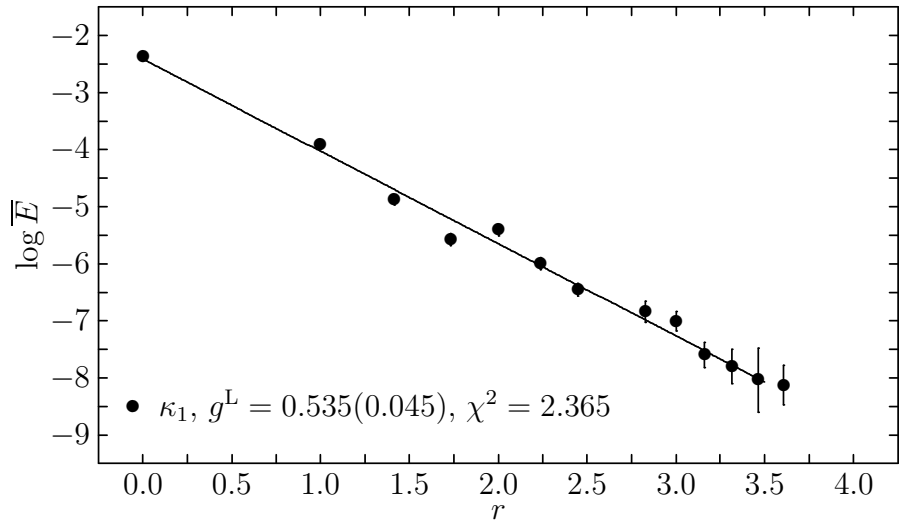
\begin{figure}
\input{fig5.tex}
\caption{Plot of $\overline{E}(r)$ as a function of
$r$ for $\kappa=0.13843$. 
\label{fig:logplot1}}
\end{figure}

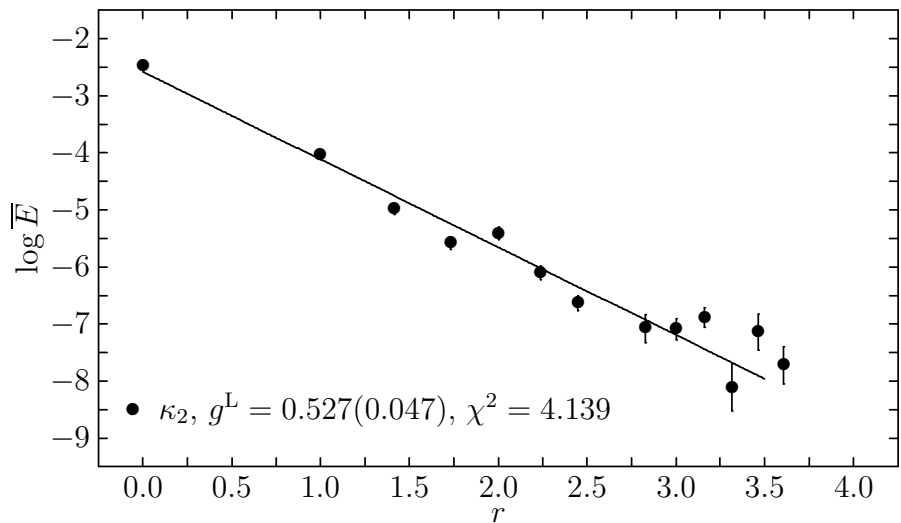
\begin{figure}
\input{fig6.tex}
\caption{Plot of $\overline{E}(r)$ as a function of
$r$ for $\kappa=0.14077$. 
\label{fig:logplot2}}
\end{figure}

Using rotational invariance, the average of the three spatial
components of $E_i(r,t)$
\begin{equation}
\overline{E}(r,t)=\frac13 (E_{1}(r,t)+ E_{2}(r,t)+ E_{3}(r,t)) 
\end{equation}
is measured. The results are reported in Figs.~\ref{figc3a}
and~\ref{figc3b}, for the two values of $\kappa$ used in this
simulation.  At fixed $r$, $\overline{E}(r,t)$ is expected to be
independent of $t$. As this behaviour is confirmed by the data, the
signal can be improved by averaging the values at different times,
each weighted with its error, yielding a function of the spatial
distance $\overline{E}(r)$, which needs to be integrated over the
three-dimensional volume. The time-slices considered in the average
are $t=4,5,6$. Logarithmic plots of $\overline{E}(r)$ is displayed in
Figs.~\ref{fig:logplot1} and~\ref{fig:logplot2} for both values of
$\kappa$, suggesting that the data are consistent with an exponential
decay.

To evaluate the volume integral, $\overline{E}(r)$ is fitted
with a two-parameter exponential for each value of $\kappa$,
\begin{equation}
f(r)=S e^{-r/r_0}
\end{equation}
The results of the fit and the values of the reduced $\chi^2$ are
recorded on the figures.  The coupling $g$ is finally obtained by
integrating analytically the fitted function:
\begin{equation}
  g^{\mathrm{L}}=-\int_0^{\infty} 4\pi r^2 \overline{E}(r) dr 
  =-\int_0^{\infty} 4\pi r^2 f(r) dr 
\end{equation}
The superscript L indicates that these numbers are defined on the
lattice using operators renormalised at the lattice energy scale
$a^{-1}=1.10$ GeV. As for the two-point functions, the value of
$g^{\mathrm{L}}$ does not appear to depend on the mass of quarks. It
is not clear from this simulation, whether this is a genuine physical
feature or, as explained earlier, an artefact of the lattice used
here. Further studies should aim to clarify this point.

The best estimate for $g^{\mathrm{L}}$ at $\kappa=\kappa_c$ is thus
obtained from a weighted average of the results at the two kappas used
in our simulations:
\begin{equation}
         g^{\mathrm{L}}=0.52(5)(10)
\label{gl}
\end{equation}
The first error is statistical; the second is a conservative estimate
of the systematic error of 20\%, based on the systematic error
observed in the estimate of $f_B$.

\section{Renormalisation constants}
\label{sec:ren}

Quantities evaluated on the lattice are connected to their continuum
counterparts by a finite renormalisation.  In order to extract
physical information, the matrix element of a bilinear quark operator
defined on the lattice, ${\cal O}^{\mathrm{L}}(a)$, has to be
multiplied by the corresponding renormalisation constant
\begin{equation}
        {\cal O}(\mu) = Z_{\cal O}(a \mu, g) \; 
        {\cal O}^{\mathrm{L}} (a)
\end{equation}
which in general depends on the renormalisation scale $\mu$ and the
details of the lattice discretization (a single continuum operator may
also match onto a set of lattice operators). For partially conserved
currents, the $\mu$ dependence disappears in the above definition, the
associated anomalous dimension being equal to zero. This is the case
for the QCD light-light axial current considered here. It is also the
case for the heavy-light current in full QCD, but not for the
static-light current onto which it matches in the effective theory:
hence the renormalisation constant $Z_A^{\mathrm{static}}$ in
Eq.~\ref{eq:ZAstatic} below, which converts the lattice matrix element
to the physical $f_B^{\mathrm{static}}$, includes the effect of
running between different scales in the effective theory.

In our simulation, we use the standard gluon action and the
tadpole-improved~\cite{lepage93}
Sheikholeslami-Wohlert~\cite{swaction} fermion action for the light
quarks:
\begin{eqnarray}
        S &=& \frac{a^4}{2\kappa} \sum_{xy}  \bar{\psi}(x) 
        \Bigg\{  \Big[ 1 - \frac{i \tilde{\kappa} a}{2 u_0^4}\sigma_{\mu \nu} 
        F_{\mu \nu}(x)  \Big] \delta_{x y} \\
        & &
        \mbox{}
        -\frac{\tilde{\kappa}}{a u_0} 
        \Big[ (1 - \gamma_\mu) U_\mu(x) \delta_{x+\hat{\mu}, y}  \nonumber
        +   (1 + \gamma_\mu) U^\dagger_\mu(y) \delta_{x-\hat{\mu}, y} \Big]
        \Bigg\}\psi (y)
\end{eqnarray}
where as usual $F_{\mu \nu}(x)$ is a lattice definition of the gluon
field strength tensor, $\tilde{\kappa} = \kappa u_0$ and we define
$u_0$ from the average plaquette in infinite volume, $u_0^4 = \langle
\frac{1}{3} \Tr U_\Box \rangle$. The redefinition of the
parameters in the action is done in order to absorb the large
renormalisation coming from lattice tadpole graphs~\cite{lepage93}.
The standard definition of $\csw$, the coefficient of the
$\sigma_{\mu\nu}F_{\mu\nu}$ term, in this case gives $\csw = 1/u_0^3$.
For our lattice, $u_0= 0.86081$ from the average plaquette, so that
$\csw = 1.57$. For the $b$ quarks we use the standard static quark
action.  Here we summarise the required renormalisation constants for
the light-light and static-light axial currents with the action and
parameters defined above.

The matrix element of the light-light axial current between an initial
state $i$ and a final state $f$ in the continuum can be written as:
\begin{equation}
        \langle f |A_\mu | i \rangle  
        =
        Z_A  \langle f | A_\mu^{\mathrm{L}} | i \rangle
        = Z_A^{\mathrm{1-loop}} \frac{u_0}{u_0^{\mathrm{1-loop}}}  
            \langle f | A_\mu^{\mathrm{L}} | i \rangle               
\end{equation}
The factor $u_0$, a measure of the average link variable, can be
interpreted as a non-perturbative rescaling of the quark fields in the
tadpole improvement prescription.  The 1-loop perturbative
renormalisation constant must then be redefined by dividing out the
1-loop expression, $u_0^{\mathrm{1-loop}}$, corresponding to the
chosen definition of $u_0$,
\begin{equation}
        Z_A^{\mathrm{tadpole}} = Z_A^{\mathrm{1-loop}}
        / u_0^{\mathrm{1-loop}}
\end{equation}
The overall renormalisation constant is given by the whole factor
\begin{equation}
Z_A = u_0 \, Z_A^{\mathrm{tadpole}}
\end{equation}
The perturbative part reads: 
\begin{equation} 
\frac{Z_A^{\mathrm{1-loop}}}{u_0^{\mathrm{1-loop}}}=
\frac{1 + \frac {\alpha C_F}{4 \pi} \zeta_A} {1 + \frac {\alpha
C_F}{4 \pi} \lambda}\simeq 1 + \frac {\alpha C_F}{4 \pi}
(\zeta_A-\lambda) 
\label{eq:zatad}
\end{equation}
where $\lambda = -\pi^2$ for our plaquette definition of $u_0$ and
$\zeta_A =-13.8$~\cite{Sachrajda,Luscher,gabrielli}.

We note here that the removal of tree-level $O(a)$ discretisation
errors is achieved by combining the Sheikholeslami-Wohlert action for
$\csw=1$ with improved operators, found by redefining or ``rotating''
the quark fields~\cite{heatlie}. In our simulation, the rotation has
not been applied to the quarks, so the perturbative coefficient
$\zeta$ above is calculated using the improved action only. Moreover,
in perturbation theory, $\csw = 1 + O(\alpha)$, so we quote $\zeta$
above calculated for $\csw=1$, without rotated light quark fields.
In~\cite{gock} the light fermion bilinear operator renormalisations
are given as functions of $\csw$ and an arbitrary amount of field
rotation: using our actual value of $\csw = 1.57$, with no field
rotation, in those results would increase $Z_A$ by $5\%$.

The mean-field improved perturbative expansion is performed in terms
of a boosted coupling constant, $\tilde\alpha$, which in our case is
chosen as $\tilde\alpha = \alpha_0/ u_0^4$, where $\alpha_0$ is the
bare lattice coupling constant. We use $\tilde\alpha$ for $\alpha$ in
Eq.~\ref{eq:zatad}. For our lattice, with $\beta=5.7$ and $u_0=
0.86081$ from the average plaquette, this leads to:
\begin{equation}
        Z_A = u_0 \, Z_A^{\mathrm{tadpole}} = 0.806.
\label{zetaa}
\end{equation}

For the heavy-light axial current computed in the static limit, the
renormalisation constant $Z_A^{\mathrm{static}}$ is found by a three
step procedure combining the matching between full QCD and the
continuum static theory at a scale of order $m_b$~\cite{eh1}, the
continuum static theory running from $m_b$ to a lattice scale
$q^*$~\cite{jimus,brogro,gim} and finally the matching between the
continuum static and the lattice static
theories~\cite{eh2,borrelli92}.  We extract from the calculation of
Borrelli and Pittori~\cite{borrelli92} the appropriate matching factor
corresponding to our case of an improved action, but without field
rotations on the light quarks. In the notation of~\cite{borrelli92}
the number we need is
\begin{equation}
    C_F \big(5/4 - A_{\gamma_0\gamma_5} - d^I -
        f^I/2 + (e-e^{\mathrm{red}})/2\big) = -19.36
\end{equation}
This appears in Eq.~\ref{eq:ZAstatic} below and contrasts with the
values $-27.2$ for the Wilson action and $-20.2$ for the improved
action with $\csw=1$ and field rotations included.

In the matching to the lattice we will adopt a Lepage-Mackenzie
choice~\cite{lepage93} of the scale $q^*$ at which to perform the
matching. This scale is determined from the expectation value of
$\ln(qa)^2$ in the one-loop lattice perturbation theory integrals for
the corrections to the renormalisation constant, including the
perturbative tadpole improvement corrections for the chosen definition
of $u_0$. For the improved, $\csw=1$, action and plaquette definition
of $u_0$, Gim\'enez and Reyes~\cite{gimrey} quote $q^* a = 2.29$. We
will use this value.

We will also adopt a plaquette definition of the lattice coupling
constant according to (for zero flavours):
\begin{equation}
     -\ln(\langle\Tr U_\Box\rangle/3) =
         {4 \pi\over3} \alpha_V(3.41/a) (1-1.19\alpha_V)
\end{equation}
Once $\alpha_V(3.41/a)$ is determined, we can use the equation for the
running of $\alpha_V$,
\begin{equation}
     \alpha_V(q) = {4 \pi \over 2 b_0 \ln(q/\Lambda_V) +
     b_1 \ln\big(2 \ln(q/\Lambda_V)\big) / b_0}
\end{equation}
to determine $\alpha_V(q^*)$. In the quenched theory, $b_0=11$ and
$b_1 = 102$. We find $a\Lambda_V = 0.294$ and $\alpha_V(q^*) = 0.216$.

Since $q^* = 2.52\gev$ is between the charm and $b$ quark thresholds
we perform the continuum running with four active flavours. From the
Particle Data Group (PDG)~\cite{pdg98}, we take
$\Lambda^{(5)}_{\overline{\mathrm{MS}}} = 237^{+26}_{-24}\mev$ using
two-loop running, and find $\Lambda^{(4)}_{\overline{\mathrm{MS}}}$
using continuity of the strong coupling at the $b$ quark
threshold. This threshold value is given by $m_b$ satisfying $m_{b \,
\overline{\mathrm{MS}}}(m_b) = m_b$. We take $m_b = 4.25\gev$, using
the average of the range, $4.1$--$4.4\gev$, quoted by the
PDG~\cite{pdg98}.

The overall renormalisation constant is thus given by:
\begin{eqnarray}
        Z_A^{\mathrm{static}} &=& 
        \bigg(\frac{\alpha(m_b)}{\alpha(q^*)}\bigg)^d
        \bigg\{1 - c \frac{\alpha(m_b)}{4 \pi}
        + \frac{\alpha(q^*) {-} \alpha(m_b)}{4 \pi} J_A
        \bigg\} \nonumber \\
        & & 
        \mbox{}
        \times \sqrt{u_0}
        \bigg(1 + \frac{\alpha_V(q^*)}{4 \pi}
        [ 4\ln(q^* a) -19.36 - C_F \lambda/2 ] \bigg)
        \label{eq:ZAstatic}
\end{eqnarray}
where $c=8/3$, $J_A=0.91$ and $d =-6/25$.  Using the inputs given
above, we find
\begin{equation}
Z_A^{\mathrm{static}}=0.78
\end{equation}
Independent variations of $\Lambda^{(5)}$ and $m_b$ to the ends of the
ranges quoted above and $\pm 10\%$ variation in $a^{-1}$ change this
value by $1.3\%$ or less. Changing $q^*$ to $1/a$ or $\pi/a$ reduces
$Z_A^{\mathrm{static}}$ by $13\%$ or raises it by $1.3\%$
respectively.
 
We close this section by noting that both $Z_A$ and
$Z_A^{\mathrm{static}}$ are evaluated in the chiral limit and so do
not depend on the light quark mass used in the simulation.

\section{Phenomenological implications}
\label{sec:pheno}

We now combine the lattice matrix element with its renormalisation
factor to find the continuum value for $g$. Since we observe no light
quark mass dependence in $g^L$ and evaluated the renormalisation
constant in the chiral limit, we multiply the lattice matrix element
in Eq.~\ref{gl} by the renormalisation constant in Eq.~\ref{zetaa}, to
obtain the value:
\begin{equation}
        g=Z_A \,
        g^{\mathrm{L}} = 0.42(4)(8)
\end{equation}
for the coupling of the heavy mesons with the Goldstone bosons.  

The direct decay $B^\ast \to B \pi$ is kinematically forbidden.
However, the corresponding reaction occurs in the $D$ system, where
the coupling $g_{D^\ast D \pi}$ is also related to $g$ by an
expression analogous to Eq.~\ref{eq:g_to_G}, although the $1/m_Q$
corrections are expected to be larger in the charm case.  A recent
analysis~\cite{iws98}, incorporating chiral symmetry breaking
corrections plus $1/m_c$ corrections in the HM$\chi$ lagrangian and
fitting to the branching ratios for $D^{*0}\to D^0\pi^0$, $D^{*+}\to
D^+\pi^0$ and $D^*_s\to D_s\pi^0$ (together with radiative decay rates
for the same $D^*$ mesons), gives
\begin{equation}
 g = \cases{0.27\errp42\errp52&{}\cr
            0.76(3)\errp21&{}\cr}
\end{equation}
The two fold ambiguity can be resolved by imposing the experimental
limit $\Gamma(D^{*+}) < 0.13\mev$~\cite{accmor92}, which gives $g <
0.52$ to a good approximation~\cite{iws98}.

Other estimates of $g$ are derived using constituent quark models and
sum rules. Starting from the non-relativistic result $g=1$, quark
models can be improved using more sophisticated assumptions to
describe the quark dynamics inside the meson. Independent estimates
are obtained from computing QCD correlation functions in the framework
of sum rules. To give an idea of the range spanned by different
determinations, some recent results are listed in
Tab.~\ref{tab:list_g}.
\begin{table}
\begin{minipage}{\linewidth}
\renewcommand{\thefootnote}{\thempfootnote}
\renewcommand{\footnoterule}{\rule{0pt}{0pt}}
\renewcommand{\footnotesep}{0pt}
\begin{tabular*}{\hsize}{@{}@{\extracolsep{\fill}}llll} 
\hline
Reference & $g$ & Reference & $g$ \\
\hline
\cite{BBKR} & 0.32(2) & \cite{nussinov87} & 0.7 \\ 
\cite{colangelo94} & 0.39 & \cite{yan92}  & 0.75-1.0 \\
\cite{casalbuoni97}\footnote{Combined sum rule + lattice results} 
& 0.44 (16) & \cite{cho92,amundson92} & 0.4-0.7 \\
  &   & \cite{colangelo94b} & 0.38 \\
\hline
\end{tabular*}
\end{minipage}
\caption{Recent determinations of the coupling constant $g$
\label{tab:list_g}}
\end{table}

The best estimate from a global analysis of available results, quoted
in the review~\cite{casalbuoni97}, is
\begin{equation}
g \simeq 0.38
\end{equation}
with a total uncertainty of 20\%; the latter being comparable with the
estimated systematic error from our lattice computation. Not only is
the result presented in this paper compatible with the previous
estimates; the systematic error is also competitive when compared to
the above results.

The agreement with other previous estimates is pleasing, but the value
of the coupling $g$ can also be used to check the consistency of the
heavy quark symmetry predictions in the soft pion limit for the
lattice results.

The form factor $f_1$ of the semileptonic decay $B \to \pi l \nu$ is
predicted to have the expression given earlier in Eq.~\ref{eq:pole}.
This behaviour for $f_1(q^2)$ is expected to be valid near zero recoil
($q^2 \to q^2_{\mathrm{max}}$), where the closest vector resonance
dominates, even beyond the leading approximation in $1/m_b$ in
HQET~\cite{Burdman94}.  One also obtains this behaviour from the heavy
meson chiral lagrangian~\cite{wise92,ccllyy,fleischer93,falkgrin94},
valid for a low momentum pion.  It is perhaps surprising that such a
behaviour is found by sum rules to hold reasonably, at least for the
$D$ meson, also at $q^2 \to 0$~\cite{BBKR}.  In this case the coupling
$g$ would fix the normalisation of the form factor $f_1$ . The
extension of the validity of the vector pole dominance for general
values of $q^2$ has no simple theoretical justification. It can be
argued that the contributions from different resonances can lead to a
single pole behaviour; however, in this case, the relation between $g$
and $f_1(0)$ would be spoilt.

Lattice data for the semi-leptonic $B$ decay form factors have been
fitted assuming a pole behaviour for $f_1$~\cite{ldd97}, yielding
$f_1(0)=0.44(3)$, to be compared with
\begin{equation}
  f_1(0)|_{\mathrm{VMD}}
   = \frac{m_B}{f_\pi f_{B^\ast}} g
\end{equation}
from Eq.~\ref{eq:pole}. Our determination of $g$ gives
$f_1(0)=0.50(5)(10)$, in good agreement with the direct fit. Such an
agreement should not come as a surprise: the lattice data, after
chiral extrapolation, turn out to be close to the end-point of the
phase space kinematically allowed for $B \to \pi$ decays ($q^2 \simeq
q^2_{max}$). The lattice normalisation of the form factor comes
therefore from a fit in a region where vector dominance does
hold. Hence, the above agreement should be seen as a consistency check
of the two lattice computations.  It is nevertheless important to see
that the two results are not contradictory.

Assuming a pole form for $f_1$, an independent bound on the value of
the residue is obtained by enforcing the theory to satisfy
unitarity~\cite{boyd94}. The bound quoted in~\cite{boyd94} is
\begin{equation}
	f_1(0) m_{B^\ast}^2 \leq 10 \gev^2
\end{equation}
which translates into
\begin{equation}
	f_1(0) \leq 0.4
\end{equation} 
This, again, agrees reasonably with the determination obtained from
VMD, using our value for $g$, and the one coming from the direct fit
of the lattice form factors. 

As the lattice determinations will improve in the future, the
comparison of the three results summarised here could become an
effective way to constrain the residue at the $B^\ast$ pole.


\section{Conclusions}

We have shown that the coupling $\gBBP$, or equivalently the coupling
$g$ in the HM$\chi$ lagrangian, may be evaluated on the lattice from
the matrix element of the light quark axial current between $B$ and
$B^\ast$ states. The value of $g$ enters many phenomenological
quantities of interest calculated in heavy meson chiral perturbation
theory, including the form factors for semileptonic $B\to\pi$ decays
mentioned here, as well as $B^\ast\to\pi$ decays and other quantities
such as ratios of heavy meson nonleptonic decay constants, heavy meson
mass splittings and radiative decays.

Even the relatively crude estimate obtained here should be interesting
for heavy meson phenomenology, but we believe future more precise
lattice computations would be valuable.

\vspace{0.8cm}

\noindent
{\bf Acknowledgements} 

It is a pleasure to acknowledge enlightening discussions with C.T.\
Sachrajda. We would also like to thank L.~Lellouch, R.~Petronzio,
G.C.\ Rossi and H.~Wittig for discussing some of the issues presented
here.

Numerical simulations were performed on a CRAY J90 at RAL under grant
GR/L55056. GdD is supported by a University of Southampton Annual
Grant Award, LDD, JMF, GdD and MDP by PPARC under grants GR/L56329 and
GR/L29927, and by EPSRC grant GR/K41663. JMF thanks the CERN Theory
Division for hospitality during the completion of this work.

\newcommand{\etal}{\emph{et al.}}

\end{document}

%% file: fig2.tex
\begin{center}                                  
                \makebox[0pt]{\begin{texdraw}
                \normalsize
                \ifx\pathDEFINED\relax\else\let\pathDEFINED\relax
                 \def\QtGfr{\ifx (\TGre \let\YhetT\cpath\else\let\YhetT\relax\fi\YhetT}
                 \def\path (#1 #2){\move (#1 #2)\futurelet\TGre\QtGfr}
                 \def\cpath (#1 #2){\lvec (#1 #2)\futurelet\TGre\QtGfr}
                \fi
                \drawdim pt
                \setunitscale 0.216 
                \linewd 3
\special{PSfile=fig2.1.ps hscale=90 vscale=90}
\move(850 25) \textref h:C v:C \htext{$t$}
\move(-20 510) \textref h:C v:C \vtext{$\log C_2^{FF}(t)$}
\textref h:C v:C
\move (232 75) \htext{3}
\move (644 75) \htext{4}
\move (1055 75) \htext{5}
\move (1467 75) \htext{6}
\textref h:R v:C
\move (135 198) \htext{$-4.0$}
\move (135 287) \htext{$-3.5$}
\move (135 376) \htext{$-3.0$}
\move (135 465) \htext{$-2.5$}
\move (135 554) \htext{$-2.0$}
\move (135 643) \htext{$-1.5$}
\move (135 732) \htext{$-1.0$}
\textref h:C v:C
\move(800 950) \textref h:C v:C \htext{}
\textref h:C v:C
\move (232 698) \htext{$\bullet$}
\textref h:C v:C
\move (644 526) \htext{$\bullet$}
\textref h:C v:C
\move (1055 359) \htext{$\bullet$}
\textref h:C v:C
\textref h:C v:C
\move (232 708) \htext{$\circ$}
\textref h:C v:C
\textref h:C v:C
\move (1055 381) \htext{$\circ$}
\textref h:C v:C
\textref h:C v:C
\move(195 855) \htext{$\bullet$}
\textref h:L v:C
\move(240 855) \htext{$\kappa_1$, $M_B=0.930(0.004)$, $Z_F^2=4.751(0.097)$, $\chi^2=1.944$}
\textref h:C v:C
\textref h:C v:C
\move(195 790) \htext{$\circ$}
\textref h:L v:C
\move(240 790) \htext{$\kappa_2$, $M_B=0.898(0.005)$, $Z_F^2=4.567(0.098)$, $\chi^2=1.594$}
\textref h:C v:C
\end{texdraw}}
\end{center}
\begin{center}                                  
                \makebox[0pt]{\begin{texdraw}
                \normalsize
                \ifx\pathDEFINED\relax\else\let\pathDEFINED\relax
                 \def\QtGfr{\ifx (\TGre \let\YhetT\cpath\else\let\YhetT\relax\fi\YhetT}
                 \def\path (#1 #2){\move (#1 #2)\futurelet\TGre\QtGfr}
                 \def\cpath (#1 #2){\lvec (#1 #2)\futurelet\TGre\QtGfr}
                \fi
                \drawdim pt
                \setunitscale 0.216 
                \linewd 3
\special{PSfile=fig2.2.ps hscale=90 vscale=90}
\move(850 25) \textref h:C v:C \htext{$t$}
\move(-20 510) \textref h:C v:C \vtext{$\log C_2^{FL}(t)$}
\textref h:C v:C
\move (232 75) \htext{3}
\move (644 75) \htext{4}
\move (1055 75) \htext{5}
\move (1467 75) \htext{6}
\textref h:R v:C
\move (135 182) \htext{$-5.0$}
\move (135 255) \htext{$-4.5$}
\move (135 328) \htext{$-4.0$}
\move (135 400) \htext{$-3.5$}
\move (135 473) \htext{$-3.0$}
\move (135 546) \htext{$-2.5$}
\move (135 619) \htext{$-2.0$}
\move (135 691) \htext{$-1.5$}
\move (135 764) \htext{$-1.0$}
\textref h:C v:C
\move(800 950) \textref h:C v:C \htext{}
\textref h:C v:C
\move (232 577) \htext{$\bullet$}
\textref h:C v:C
\move (644 441) \htext{$\bullet$}
\textref h:C v:C
\move (1055 305) \htext{$\bullet$}
\textref h:C v:C
\move (1467 170) \htext{$\bullet$}
\textref h:C v:C
\move (232 585) \htext{$\circ$}
\textref h:C v:C
\move (644 452) \htext{$\circ$}
\textref h:C v:C
\move (1055 320) \htext{$\circ$}
\textref h:C v:C
\move (1467 189) \htext{$\circ$}
\textref h:C v:C
\move(195 855) \htext{$\bullet$}
\textref h:L v:C
\move(240 855) \htext{$\kappa_1$, $M_B=0.934(0.002)$, $Z_L^2=0.591(0.014)$, $\chi^2=0.742$}
\textref h:C v:C
\textref h:C v:C
\move(195 790) \htext{$\circ$}
\textref h:L v:C
\move(240 790) \htext{$\kappa_2$, $M_B=0.908(0.002)$, $Z_L^2=0.583(0.015)$, $\chi^2=1.546$}
\textref h:C v:C
\end{texdraw}}
\end{center}

%% file: fig3.tex
\begin{center}                                  
                \makebox[0pt]{\begin{texdraw}
                \normalsize
                \ifx\pathDEFINED\relax\else\let\pathDEFINED\relax
                 \def\QtGfr{\ifx (\TGre \let\YhetT\cpath\else\let\YhetT\relax\fi\YhetT}
                 \def\path (#1 #2){\move (#1 #2)\futurelet\TGre\QtGfr}
                 \def\cpath (#1 #2){\lvec (#1 #2)\futurelet\TGre\QtGfr}
                \fi
                \drawdim pt
                \setunitscale 0.216 
                \linewd 3
\special{PSfile=fig3.1.ps hscale=90 vscale=90}
\move(850 25) \textref h:C v:C \htext{distance $r$}
\move(25 510) \textref h:C v:C \vtext{}
\linewd 3
\textref h:C v:C
\move (266 75) \htext{0}
\move (500 75) \htext{2}
\move (733 75) \htext{4}
\move (966 75) \htext{6}
\move (1200 75) \htext{8}
\move (1433 75) \htext{10}
\textref h:R v:C
\move (135 243) \htext{$-0.020$}
\move (135 376) \htext{$-0.015$}
\move (135 509) \htext{$-0.010$}
\move (135 643) \htext{$-0.005$}
\move (135 776) \htext{$0.000$}
\textref h:C v:C
\move(800 950) \textref h:C v:C \htext{$E^0(r,t), \kappa=0.13843$}
\textref h:C v:C
\move (266 405) \htext{$\bullet$}
\textref h:C v:C
\move (383 604) \htext{$\bullet$}
\textref h:C v:C
\move (431 682) \htext{$\bullet$}
\textref h:C v:C
\move (468 723) \htext{$\bullet$}
\textref h:C v:C
\move (500 730) \htext{$\bullet$}
\textref h:C v:C
\move (527 746) \htext{$\bullet$}
\textref h:C v:C
\move (552 756) \htext{$\bullet$}
\textref h:C v:C
\move (596 763) \htext{$\bullet$}
\textref h:C v:C
\move (616 766) \htext{$\bullet$}
\textref h:C v:C
\move (635 767) \htext{$\bullet$}
\textref h:C v:C
\move (653 770) \htext{$\bullet$}
\textref h:C v:C
\move (670 770) \htext{$\bullet$}
\textref h:C v:C
\move (687 771) \htext{$\bullet$}
\textref h:C v:C
\move (703 773) \htext{$\bullet$}
\textref h:C v:C
\move (733 773) \htext{$\bullet$}
\textref h:C v:C
\move (747 774) \htext{$\bullet$}
\textref h:C v:C
\move (761 774) \htext{$\bullet$}
\textref h:C v:C
\move (775 774) \htext{$\bullet$}
\textref h:C v:C
\move (788 775) \htext{$\bullet$}
\textref h:C v:C
\move (801 775) \htext{$\bullet$}
\textref h:C v:C
\move (813 776) \htext{$\bullet$}
\textref h:C v:C
\move (838 775) \htext{$\bullet$}
\textref h:C v:C
\move (850 776) \htext{$\bullet$}
\textref h:C v:C
\move (861 776) \htext{$\bullet$}
\textref h:C v:C
\move (872 775) \htext{$\bullet$}
\textref h:C v:C
\move (894 776) \htext{$\bullet$}
\textref h:C v:C
\move (905 774) \htext{$\bullet$}
\textref h:C v:C
\move (926 775) \htext{$\bullet$}
\textref h:C v:C
\move (936 775) \htext{$\bullet$}
\textref h:C v:C
\move (946 776) \htext{$\bullet$}
\textref h:C v:C
\move (956 776) \htext{$\bullet$}
\textref h:C v:C
\move (966 776) \htext{$\bullet$}
\textref h:C v:C
\move (976 774) \htext{$\bullet$}
\textref h:C v:C
\move (985 776) \htext{$\bullet$}
\textref h:C v:C
\move (1004 773) \htext{$\bullet$}
\textref h:C v:C
\move (1013 775) \htext{$\bullet$}
\textref h:C v:C
\move (1022 775) \htext{$\bullet$}
\textref h:C v:C
\move (1031 776) \htext{$\bullet$}
\textref h:C v:C
\move (1040 775) \htext{$\bullet$}
\textref h:C v:C
\move (1049 777) \htext{$\bullet$}
\textref h:C v:C
\move (1057 777) \htext{$\bullet$}
\textref h:C v:C
\move (1074 774) \htext{$\bullet$}
\textref h:C v:C
\move (1083 775) \htext{$\bullet$}
\textref h:C v:C
\move (1091 776) \htext{$\bullet$}
\textref h:C v:C
\move (1099 776) \htext{$\bullet$}
\textref h:C v:C
\move (1107 773) \htext{$\bullet$}
\textref h:C v:C
\move (1116 775) \htext{$\bullet$}
\textref h:C v:C
\move (1123 776) \htext{$\bullet$}
\textref h:C v:C
\move (1139 776) \htext{$\bullet$}
\textref h:C v:C
\move (1147 775) \htext{$\bullet$}
\textref h:C v:C
\move (1162 775) \htext{$\bullet$}
\textref h:C v:C
\move (1177 777) \htext{$\bullet$}
\textref h:C v:C
\move (1185 777) \htext{$\bullet$}
\textref h:C v:C
\move (1207 775) \htext{$\bullet$}
\textref h:C v:C
\move (1214 775) \htext{$\bullet$}
\textref h:C v:C
\move (1228 775) \htext{$\bullet$}
\textref h:C v:C
\move (1242 775) \htext{$\bullet$}
\textref h:C v:C
\move (1256 780) \htext{$\bullet$}
\textref h:C v:C
\move (1263 776) \htext{$\bullet$}
\textref h:C v:C
\move (1277 773) \htext{$\bullet$}
\textref h:C v:C
\move (1283 776) \htext{$\bullet$}
\textref h:C v:C
\move (1290 776) \htext{$\bullet$}
\textref h:C v:C
\move (1316 774) \htext{$\bullet$}
\textref h:C v:C
\move (1348 775) \htext{$\bullet$}
\textref h:C v:C
\move (1361 773) \htext{$\bullet$}
\textref h:C v:C
\move (1415 776) \htext{$\bullet$}
\textref h:C v:C
\move (1479 775) \htext{$\bullet$}
\textref h:C v:C
\move(650 410) \htext{$\bullet$}
\textref h:L v:C
\move(695 410) \htext{$t=3$
}
\textref h:C v:C
\textref h:C v:C
\move (266 506) \htext{$\Diamond$}
\textref h:C v:C
\move (383 648) \htext{$\Diamond$}
\textref h:C v:C
\move (431 704) \htext{$\Diamond$}
\textref h:C v:C
\move (468 735) \htext{$\Diamond$}
\textref h:C v:C
\move (500 739) \htext{$\Diamond$}
\textref h:C v:C
\move (527 748) \htext{$\Diamond$}
\textref h:C v:C
\move (552 759) \htext{$\Diamond$}
\textref h:C v:C
\move (596 763) \htext{$\Diamond$}
\textref h:C v:C
\move (616 764) \htext{$\Diamond$}
\textref h:C v:C
\move (635 765) \htext{$\Diamond$}
\textref h:C v:C
\move (653 770) \htext{$\Diamond$}
\textref h:C v:C
\move (670 773) \htext{$\Diamond$}
\textref h:C v:C
\move (687 771) \htext{$\Diamond$}
\textref h:C v:C
\move (703 772) \htext{$\Diamond$}
\textref h:C v:C
\move (733 774) \htext{$\Diamond$}
\textref h:C v:C
\move (747 773) \htext{$\Diamond$}
\textref h:C v:C
\move (761 774) \htext{$\Diamond$}
\textref h:C v:C
\move (775 775) \htext{$\Diamond$}
\textref h:C v:C
\move (788 775) \htext{$\Diamond$}
\textref h:C v:C
\move (801 774) \htext{$\Diamond$}
\textref h:C v:C
\move (813 771) \htext{$\Diamond$}
\textref h:C v:C
\move (838 773) \htext{$\Diamond$}
\textref h:C v:C
\move (850 774) \htext{$\Diamond$}
\textref h:C v:C
\move (861 774) \htext{$\Diamond$}
\textref h:C v:C
\move (872 777) \htext{$\Diamond$}
\textref h:C v:C
\move (894 774) \htext{$\Diamond$}
\textref h:C v:C
\move (905 777) \htext{$\Diamond$}
\textref h:C v:C
\move (926 773) \htext{$\Diamond$}
\textref h:C v:C
\move (936 773) \htext{$\Diamond$}
\textref h:C v:C
\move (946 775) \htext{$\Diamond$}
\textref h:C v:C
\move (956 776) \htext{$\Diamond$}
\textref h:C v:C
\move (966 775) \htext{$\Diamond$}
\textref h:C v:C
\move (976 771) \htext{$\Diamond$}
\textref h:C v:C
\move (985 773) \htext{$\Diamond$}
\textref h:C v:C
\move (1004 771) \htext{$\Diamond$}
\textref h:C v:C
\move (1013 774) \htext{$\Diamond$}
\textref h:C v:C
\move (1022 776) \htext{$\Diamond$}
\textref h:C v:C
\move (1031 774) \htext{$\Diamond$}
\textref h:C v:C
\move (1040 776) \htext{$\Diamond$}
\textref h:C v:C
\move (1049 775) \htext{$\Diamond$}
\textref h:C v:C
\move (1057 775) \htext{$\Diamond$}
\textref h:C v:C
\move (1074 780) \htext{$\Diamond$}
\textref h:C v:C
\move (1083 773) \htext{$\Diamond$}
\textref h:C v:C
\move (1091 774) \htext{$\Diamond$}
\textref h:C v:C
\move (1099 772) \htext{$\Diamond$}
\textref h:C v:C
\move (1107 769) \htext{$\Diamond$}
\textref h:C v:C
\move (1116 773) \htext{$\Diamond$}
\textref h:C v:C
\move (1123 775) \htext{$\Diamond$}
\textref h:C v:C
\move (1139 773) \htext{$\Diamond$}
\textref h:C v:C
\move (1147 774) \htext{$\Diamond$}
\textref h:C v:C
\move (1162 775) \htext{$\Diamond$}
\textref h:C v:C
\move (1177 775) \htext{$\Diamond$}
\textref h:C v:C
\move (1185 775) \htext{$\Diamond$}
\textref h:C v:C
\move (1207 773) \htext{$\Diamond$}
\textref h:C v:C
\move (1214 775) \htext{$\Diamond$}
\textref h:C v:C
\move (1228 775) \htext{$\Diamond$}
\textref h:C v:C
\move (1242 779) \htext{$\Diamond$}
\textref h:C v:C
\move (1256 772) \htext{$\Diamond$}
\textref h:C v:C
\move (1263 781) \htext{$\Diamond$}
\textref h:C v:C
\move (1277 773) \htext{$\Diamond$}
\textref h:C v:C
\move (1283 775) \htext{$\Diamond$}
\textref h:C v:C
\move (1290 776) \htext{$\Diamond$}
\textref h:C v:C
\move (1316 777) \htext{$\Diamond$}
\textref h:C v:C
\move (1348 771) \htext{$\Diamond$}
\textref h:C v:C
\move (1361 769) \htext{$\Diamond$}
\textref h:C v:C
\move (1415 773) \htext{$\Diamond$}
\textref h:C v:C
\move (1479 772) \htext{$\Diamond$}
\textref h:C v:C
\move(650 365) \htext{$\Diamond$}
\textref h:L v:C
\move(695 365) \htext{$t=4$
}
\textref h:C v:C
\textref h:C v:C
\move (266 589) \htext{$\circ$}
\textref h:C v:C
\move (383 704) \htext{$\circ$}
\textref h:C v:C
\move (431 731) \htext{$\circ$}
\textref h:C v:C
\move (468 754) \htext{$\circ$}
\textref h:C v:C
\move (500 767) \htext{$\circ$}
\textref h:C v:C
\move (527 757) \htext{$\circ$}
\textref h:C v:C
\move (552 763) \htext{$\circ$}
\textref h:C v:C
\move (596 759) \htext{$\circ$}
\textref h:C v:C
\move (616 763) \htext{$\circ$}
\textref h:C v:C
\move (635 772) \htext{$\circ$}
\textref h:C v:C
\move (653 773) \htext{$\circ$}
\textref h:C v:C
\move (670 776) \htext{$\circ$}
\textref h:C v:C
\move (687 778) \htext{$\circ$}
\textref h:C v:C
\move (703 770) \htext{$\circ$}
\textref h:C v:C
\move (733 781) \htext{$\circ$}
\textref h:C v:C
\move (747 778) \htext{$\circ$}
\textref h:C v:C
\move (761 773) \htext{$\circ$}
\textref h:C v:C
\move (775 769) \htext{$\circ$}
\textref h:C v:C
\move (788 772) \htext{$\circ$}
\textref h:C v:C
\move (801 777) \htext{$\circ$}
\textref h:C v:C
\move (813 766) \htext{$\circ$}
\textref h:C v:C
\move (838 769) \htext{$\circ$}
\textref h:C v:C
\move (850 773) \htext{$\circ$}
\textref h:C v:C
\move (861 773) \htext{$\circ$}
\textref h:C v:C
\move (872 774) \htext{$\circ$}
\textref h:C v:C
\move (894 772) \htext{$\circ$}
\textref h:C v:C
\move (905 773) \htext{$\circ$}
\textref h:C v:C
\move (926 780) \htext{$\circ$}
\textref h:C v:C
\move (936 775) \htext{$\circ$}
\textref h:C v:C
\move (946 778) \htext{$\circ$}
\textref h:C v:C
\move (956 774) \htext{$\circ$}
\textref h:C v:C
\move (966 771) \htext{$\circ$}
\textref h:C v:C
\move (976 756) \htext{$\circ$}
\textref h:C v:C
\move (985 765) \htext{$\circ$}
\textref h:C v:C
\move (1004 761) \htext{$\circ$}
\textref h:C v:C
\move (1013 775) \htext{$\circ$}
\textref h:C v:C
\move (1022 779) \htext{$\circ$}
\textref h:C v:C
\move (1031 778) \htext{$\circ$}
\textref h:C v:C
\move (1040 787) \htext{$\circ$}
\textref h:C v:C
\move (1049 780) \htext{$\circ$}
\textref h:C v:C
\move (1057 778) \htext{$\circ$}
\textref h:C v:C
\move (1074 781) \htext{$\circ$}
\textref h:C v:C
\move (1083 764) \htext{$\circ$}
\textref h:C v:C
\move (1091 771) \htext{$\circ$}
\textref h:C v:C
\move (1099 765) \htext{$\circ$}
\textref h:C v:C
\move (1107 774) \htext{$\circ$}
\textref h:C v:C
\move (1116 775) \htext{$\circ$}
\textref h:C v:C
\move (1123 772) \htext{$\circ$}
\textref h:C v:C
\move (1139 768) \htext{$\circ$}
\textref h:C v:C
\move (1147 763) \htext{$\circ$}
\textref h:C v:C
\move (1162 773) \htext{$\circ$}
\textref h:C v:C
\move (1177 765) \htext{$\circ$}
\textref h:C v:C
\move (1185 767) \htext{$\circ$}
\textref h:C v:C
\move (1207 763) \htext{$\circ$}
\textref h:C v:C
\move (1214 768) \htext{$\circ$}
\textref h:C v:C
\move (1228 760) \htext{$\circ$}
\textref h:C v:C
\move (1242 776) \htext{$\circ$}
\textref h:C v:C
\move (1256 797) \htext{$\circ$}
\textref h:C v:C
\move (1263 799) \htext{$\circ$}
\textref h:C v:C
\move (1277 785) \htext{$\circ$}
\textref h:C v:C
\move (1283 767) \htext{$\circ$}
\textref h:C v:C
\move (1290 770) \htext{$\circ$}
\textref h:C v:C
\move (1316 756) \htext{$\circ$}
\textref h:C v:C
\move (1348 773) \htext{$\circ$}
\textref h:C v:C
\move (1361 764) \htext{$\circ$}
\textref h:C v:C
\move (1415 776) \htext{$\circ$}
\textref h:C v:C
\move (1479 783) \htext{$\circ$}
\textref h:C v:C
\move(650 320) \htext{$\circ$}
\textref h:L v:C
\move(695 320) \htext{$t=5$
}
\textref h:C v:C
\textref h:C v:C
\move (266 812) \htext{$\Box$}
\textref h:C v:C
\move (383 764) \htext{$\Box$}
\textref h:C v:C
\move (431 763) \htext{$\Box$}
\textref h:C v:C
\move (468 722) \htext{$\Box$}
\textref h:C v:C
\move (500 750) \htext{$\Box$}
\textref h:C v:C
\move (527 769) \htext{$\Box$}
\textref h:C v:C
\move (552 761) \htext{$\Box$}
\textref h:C v:C
\move (596 757) \htext{$\Box$}
\textref h:C v:C
\move (616 729) \htext{$\Box$}
\textref h:C v:C
\move (635 783) \htext{$\Box$}
\textref h:C v:C
\move (653 755) \htext{$\Box$}
\textref h:C v:C
\move (670 758) \htext{$\Box$}
\textref h:C v:C
\move (687 802) \htext{$\Box$}
\textref h:C v:C
\move (703 773) \htext{$\Box$}
\textref h:C v:C
\move (733 778) \htext{$\Box$}
\textref h:C v:C
\move (747 765) \htext{$\Box$}
\textref h:C v:C
\move (761 766) \htext{$\Box$}
\textref h:C v:C
\move (775 754) \htext{$\Box$}
\textref h:C v:C
\move (788 765) \htext{$\Box$}
\textref h:C v:C
\move (801 758) \htext{$\Box$}
\textref h:C v:C
\move (813 813) \htext{$\Box$}
\textref h:C v:C
\move (838 739) \htext{$\Box$}
\textref h:C v:C
\move (850 796) \htext{$\Box$}
\textref h:C v:C
\move (861 773) \htext{$\Box$}
\textref h:C v:C
\move (872 733) \htext{$\Box$}
\textref h:C v:C
\move (894 777) \htext{$\Box$}
\textref h:C v:C
\move (905 761) \htext{$\Box$}
\textref h:C v:C
\move (926 807) \htext{$\Box$}
\textref h:C v:C
\move (936 790) \htext{$\Box$}
\textref h:C v:C
\move (946 764) \htext{$\Box$}
\textref h:C v:C
\move (956 776) \htext{$\Box$}
\textref h:C v:C
\move (966 754) \htext{$\Box$}
\textref h:C v:C
\move (976 732) \htext{$\Box$}
\textref h:C v:C
\move (985 752) \htext{$\Box$}
\textref h:C v:C
\move (1004 788) \htext{$\Box$}
\textref h:C v:C
\move (1013 778) \htext{$\Box$}
\textref h:C v:C
\move (1022 804) \htext{$\Box$}
\textref h:C v:C
\move (1031 758) \htext{$\Box$}
\textref h:C v:C
\move (1040 788) \htext{$\Box$}
\textref h:C v:C
\move (1049 777) \htext{$\Box$}
\textref h:C v:C
\move (1057 780) \htext{$\Box$}
\textref h:C v:C
\move (1074 736) \htext{$\Box$}
\textref h:C v:C
\move (1083 726) \htext{$\Box$}
\textref h:C v:C
\move (1091 770) \htext{$\Box$}
\textref h:C v:C
\move (1099 756) \htext{$\Box$}
\textref h:C v:C
\move (1107 731) \htext{$\Box$}
\textref h:C v:C
\move (1116 771) \htext{$\Box$}
\textref h:C v:C
\move (1123 800) \htext{$\Box$}
\textref h:C v:C
\move (1139 773) \htext{$\Box$}
\textref h:C v:C
\move (1147 736) \htext{$\Box$}
\textref h:C v:C
\move (1162 772) \htext{$\Box$}
\textref h:C v:C
\move (1177 778) \htext{$\Box$}
\textref h:C v:C
\move (1185 770) \htext{$\Box$}
\textref h:C v:C
\move (1207 757) \htext{$\Box$}
\textref h:C v:C
\move (1214 722) \htext{$\Box$}
\textref h:C v:C
\move (1228 646) \htext{$\Box$}
\textref h:C v:C
\move (1242 769) \htext{$\Box$}
\textref h:C v:C
\move (1256 820) \htext{$\Box$}
\textref h:C v:C
\move (1263 848) \htext{$\Box$}
\textref h:C v:C
\move (1277 745) \htext{$\Box$}
\textref h:C v:C
\move (1283 713) \htext{$\Box$}
\textref h:C v:C
\move (1290 776) \htext{$\Box$}
\textref h:C v:C
\move (1316 787) \htext{$\Box$}
\textref h:C v:C
\move (1348 729) \htext{$\Box$}
\textref h:C v:C
\move (1361 761) \htext{$\Box$}
\textref h:C v:C
\move (1415 768) \htext{$\Box$}
\textref h:C v:C
\textref h:C v:C
\move(650 275) \htext{$\Box$}
\textref h:L v:C
\move(695 275) \htext{$t=6$
}
\textref h:C v:C
\end{texdraw}}
\end{center}
\vspace{1em}
\begin{center}                                  
                \makebox[0pt]{\begin{texdraw}
                \normalsize
                \ifx\pathDEFINED\relax\else\let\pathDEFINED\relax
                 \def\QtGfr{\ifx (\TGre \let\YhetT\cpath\else\let\YhetT\relax\fi\YhetT}
                 \def\path (#1 #2){\move (#1 #2)\futurelet\TGre\QtGfr}
                 \def\cpath (#1 #2){\lvec (#1 #2)\futurelet\TGre\QtGfr}
                \fi
                \drawdim pt
                \setunitscale 0.216 
                \linewd 3
\special{PSfile=fig3.2.ps hscale=90 vscale=90}
\move(850 25) \textref h:C v:C \htext{distance $r$}
\move(25 510) \textref h:C v:C \vtext{}
\linewd 3
\textref h:C v:C
\move (266 75) \htext{0}
\move (500 75) \htext{2}
\move (733 75) \htext{4}
\move (966 75) \htext{6}
\move (1200 75) \htext{8}
\move (1433 75) \htext{10}
\textref h:R v:C
\move (135 224) \htext{$-0.10$}
\move (135 338) \htext{$-0.08$}
\move (135 452) \htext{$-0.06$}
\move (135 567) \htext{$-0.04$}
\move (135 681) \htext{$-0.02$}
\move (135 795) \htext{$0.00$}
\textref h:C v:C
\move(800 950) \textref h:C v:C \htext{$\overline{E}(r,t), \kappa=0.13843$}
\textref h:C v:C
\move (266 300) \htext{$\bullet$}
\textref h:C v:C
\move (383 683) \htext{$\bullet$}
\textref h:C v:C
\move (431 758) \htext{$\bullet$}
\textref h:C v:C
\move (468 780) \htext{$\bullet$}
\textref h:C v:C
\move (500 772) \htext{$\bullet$}
\textref h:C v:C
\move (527 785) \htext{$\bullet$}
\textref h:C v:C
\move (552 789) \htext{$\bullet$}
\textref h:C v:C
\move (596 791) \htext{$\bullet$}
\textref h:C v:C
\move (616 792) \htext{$\bullet$}
\textref h:C v:C
\move (635 792) \htext{$\bullet$}
\textref h:C v:C
\move (653 793) \htext{$\bullet$}
\textref h:C v:C
\move (670 794) \htext{$\bullet$}
\textref h:C v:C
\move (687 794) \htext{$\bullet$}
\textref h:C v:C
\move (703 794) \htext{$\bullet$}
\textref h:C v:C
\move (733 794) \htext{$\bullet$}
\textref h:C v:C
\move (747 795) \htext{$\bullet$}
\textref h:C v:C
\move (761 794) \htext{$\bullet$}
\textref h:C v:C
\move (775 794) \htext{$\bullet$}
\textref h:C v:C
\move (788 795) \htext{$\bullet$}
\textref h:C v:C
\move (801 795) \htext{$\bullet$}
\textref h:C v:C
\move (813 795) \htext{$\bullet$}
\textref h:C v:C
\move (838 795) \htext{$\bullet$}
\textref h:C v:C
\move (850 795) \htext{$\bullet$}
\textref h:C v:C
\move (861 795) \htext{$\bullet$}
\textref h:C v:C
\move (872 795) \htext{$\bullet$}
\textref h:C v:C
\move (894 795) \htext{$\bullet$}
\textref h:C v:C
\move (905 795) \htext{$\bullet$}
\textref h:C v:C
\move (926 795) \htext{$\bullet$}
\textref h:C v:C
\move (936 795) \htext{$\bullet$}
\textref h:C v:C
\move (946 795) \htext{$\bullet$}
\textref h:C v:C
\move (956 795) \htext{$\bullet$}
\textref h:C v:C
\move (966 795) \htext{$\bullet$}
\textref h:C v:C
\move (976 795) \htext{$\bullet$}
\textref h:C v:C
\move (985 795) \htext{$\bullet$}
\textref h:C v:C
\move (1004 795) \htext{$\bullet$}
\textref h:C v:C
\move (1013 795) \htext{$\bullet$}
\textref h:C v:C
\move (1022 795) \htext{$\bullet$}
\textref h:C v:C
\move (1031 795) \htext{$\bullet$}
\textref h:C v:C
\move (1040 796) \htext{$\bullet$}
\textref h:C v:C
\move (1049 795) \htext{$\bullet$}
\textref h:C v:C
\move (1057 795) \htext{$\bullet$}
\textref h:C v:C
\move (1074 795) \htext{$\bullet$}
\textref h:C v:C
\move (1083 795) \htext{$\bullet$}
\textref h:C v:C
\move (1091 795) \htext{$\bullet$}
\textref h:C v:C
\move (1099 795) \htext{$\bullet$}
\textref h:C v:C
\move (1107 796) \htext{$\bullet$}
\textref h:C v:C
\move (1116 795) \htext{$\bullet$}
\textref h:C v:C
\move (1123 795) \htext{$\bullet$}
\textref h:C v:C
\move (1139 795) \htext{$\bullet$}
\textref h:C v:C
\move (1147 795) \htext{$\bullet$}
\textref h:C v:C
\move (1162 795) \htext{$\bullet$}
\textref h:C v:C
\move (1177 795) \htext{$\bullet$}
\textref h:C v:C
\move (1185 795) \htext{$\bullet$}
\textref h:C v:C
\move (1207 795) \htext{$\bullet$}
\textref h:C v:C
\move (1214 795) \htext{$\bullet$}
\textref h:C v:C
\move (1228 795) \htext{$\bullet$}
\textref h:C v:C
\move (1242 795) \htext{$\bullet$}
\textref h:C v:C
\move (1256 795) \htext{$\bullet$}
\textref h:C v:C
\move (1263 795) \htext{$\bullet$}
\textref h:C v:C
\move (1277 795) \htext{$\bullet$}
\textref h:C v:C
\move (1283 795) \htext{$\bullet$}
\textref h:C v:C
\move (1290 795) \htext{$\bullet$}
\textref h:C v:C
\move (1316 795) \htext{$\bullet$}
\textref h:C v:C
\move (1348 794) \htext{$\bullet$}
\textref h:C v:C
\move (1361 795) \htext{$\bullet$}
\textref h:C v:C
\move (1415 794) \htext{$\bullet$}
\textref h:C v:C
\move (1479 794) \htext{$\bullet$}
\textref h:C v:C
\move(650 410) \htext{$\bullet$}
\textref h:L v:C
\move(695 410) \htext{$t=3$
}
\textref h:C v:C
\textref h:C v:C
\move (266 268) \htext{$\Diamond$}
\textref h:C v:C
\move (383 683) \htext{$\Diamond$}
\textref h:C v:C
\move (431 753) \htext{$\Diamond$}
\textref h:C v:C
\move (468 774) \htext{$\Diamond$}
\textref h:C v:C
\move (500 768) \htext{$\Diamond$}
\textref h:C v:C
\move (527 782) \htext{$\Diamond$}
\textref h:C v:C
\move (552 786) \htext{$\Diamond$}
\textref h:C v:C
\move (596 789) \htext{$\Diamond$}
\textref h:C v:C
\move (616 790) \htext{$\Diamond$}
\textref h:C v:C
\move (635 792) \htext{$\Diamond$}
\textref h:C v:C
\move (653 793) \htext{$\Diamond$}
\textref h:C v:C
\move (670 793) \htext{$\Diamond$}
\textref h:C v:C
\move (687 794) \htext{$\Diamond$}
\textref h:C v:C
\move (703 793) \htext{$\Diamond$}
\textref h:C v:C
\move (733 793) \htext{$\Diamond$}
\textref h:C v:C
\move (747 795) \htext{$\Diamond$}
\textref h:C v:C
\move (761 794) \htext{$\Diamond$}
\textref h:C v:C
\move (775 795) \htext{$\Diamond$}
\textref h:C v:C
\move (788 794) \htext{$\Diamond$}
\textref h:C v:C
\move (801 795) \htext{$\Diamond$}
\textref h:C v:C
\move (813 794) \htext{$\Diamond$}
\textref h:C v:C
\move (838 795) \htext{$\Diamond$}
\textref h:C v:C
\move (850 794) \htext{$\Diamond$}
\textref h:C v:C
\move (861 795) \htext{$\Diamond$}
\textref h:C v:C
\move (872 795) \htext{$\Diamond$}
\textref h:C v:C
\move (894 795) \htext{$\Diamond$}
\textref h:C v:C
\move (905 795) \htext{$\Diamond$}
\textref h:C v:C
\move (926 796) \htext{$\Diamond$}
\textref h:C v:C
\move (936 795) \htext{$\Diamond$}
\textref h:C v:C
\move (946 795) \htext{$\Diamond$}
\textref h:C v:C
\move (956 795) \htext{$\Diamond$}
\textref h:C v:C
\move (966 795) \htext{$\Diamond$}
\textref h:C v:C
\move (976 795) \htext{$\Diamond$}
\textref h:C v:C
\move (985 796) \htext{$\Diamond$}
\textref h:C v:C
\move (1004 795) \htext{$\Diamond$}
\textref h:C v:C
\move (1013 795) \htext{$\Diamond$}
\textref h:C v:C
\move (1022 795) \htext{$\Diamond$}
\textref h:C v:C
\move (1031 796) \htext{$\Diamond$}
\textref h:C v:C
\move (1040 796) \htext{$\Diamond$}
\textref h:C v:C
\move (1049 794) \htext{$\Diamond$}
\textref h:C v:C
\move (1057 795) \htext{$\Diamond$}
\textref h:C v:C
\move (1074 795) \htext{$\Diamond$}
\textref h:C v:C
\move (1083 796) \htext{$\Diamond$}
\textref h:C v:C
\move (1091 794) \htext{$\Diamond$}
\textref h:C v:C
\move (1099 795) \htext{$\Diamond$}
\textref h:C v:C
\move (1107 794) \htext{$\Diamond$}
\textref h:C v:C
\move (1116 795) \htext{$\Diamond$}
\textref h:C v:C
\move (1123 795) \htext{$\Diamond$}
\textref h:C v:C
\move (1139 795) \htext{$\Diamond$}
\textref h:C v:C
\move (1147 794) \htext{$\Diamond$}
\textref h:C v:C
\move (1162 794) \htext{$\Diamond$}
\textref h:C v:C
\move (1177 794) \htext{$\Diamond$}
\textref h:C v:C
\move (1185 796) \htext{$\Diamond$}
\textref h:C v:C
\move (1207 795) \htext{$\Diamond$}
\textref h:C v:C
\move (1214 796) \htext{$\Diamond$}
\textref h:C v:C
\move (1228 794) \htext{$\Diamond$}
\textref h:C v:C
\move (1242 794) \htext{$\Diamond$}
\textref h:C v:C
\move (1256 793) \htext{$\Diamond$}
\textref h:C v:C
\move (1263 796) \htext{$\Diamond$}
\textref h:C v:C
\move (1277 795) \htext{$\Diamond$}
\textref h:C v:C
\move (1283 796) \htext{$\Diamond$}
\textref h:C v:C
\move (1290 796) \htext{$\Diamond$}
\textref h:C v:C
\move (1316 792) \htext{$\Diamond$}
\textref h:C v:C
\move (1348 794) \htext{$\Diamond$}
\textref h:C v:C
\move (1361 795) \htext{$\Diamond$}
\textref h:C v:C
\move (1415 794) \htext{$\Diamond$}
\textref h:C v:C
\move (1479 791) \htext{$\Diamond$}
\textref h:C v:C
\move(650 365) \htext{$\Diamond$}
\textref h:L v:C
\move(695 365) \htext{$t=4$
}
\textref h:C v:C
\textref h:C v:C
\move (266 254) \htext{$\circ$}
\textref h:C v:C
\move (383 680) \htext{$\circ$}
\textref h:C v:C
\move (431 750) \htext{$\circ$}
\textref h:C v:C
\move (468 772) \htext{$\circ$}
\textref h:C v:C
\move (500 772) \htext{$\circ$}
\textref h:C v:C
\move (527 778) \htext{$\circ$}
\textref h:C v:C
\move (552 786) \htext{$\circ$}
\textref h:C v:C
\move (596 787) \htext{$\circ$}
\textref h:C v:C
\move (616 790) \htext{$\circ$}
\textref h:C v:C
\move (635 793) \htext{$\circ$}
\textref h:C v:C
\move (653 794) \htext{$\circ$}
\textref h:C v:C
\move (670 793) \htext{$\circ$}
\textref h:C v:C
\move (687 793) \htext{$\circ$}
\textref h:C v:C
\move (703 793) \htext{$\circ$}
\textref h:C v:C
\move (733 798) \htext{$\circ$}
\textref h:C v:C
\move (747 795) \htext{$\circ$}
\textref h:C v:C
\move (761 793) \htext{$\circ$}
\textref h:C v:C
\move (775 797) \htext{$\circ$}
\textref h:C v:C
\move (788 790) \htext{$\circ$}
\textref h:C v:C
\move (801 796) \htext{$\circ$}
\textref h:C v:C
\move (813 798) \htext{$\circ$}
\textref h:C v:C
\move (838 797) \htext{$\circ$}
\textref h:C v:C
\move (850 794) \htext{$\circ$}
\textref h:C v:C
\move (861 796) \htext{$\circ$}
\textref h:C v:C
\move (872 794) \htext{$\circ$}
\textref h:C v:C
\move (894 796) \htext{$\circ$}
\textref h:C v:C
\move (905 793) \htext{$\circ$}
\textref h:C v:C
\move (926 797) \htext{$\circ$}
\textref h:C v:C
\move (936 794) \htext{$\circ$}
\textref h:C v:C
\move (946 796) \htext{$\circ$}
\textref h:C v:C
\move (956 795) \htext{$\circ$}
\textref h:C v:C
\move (966 797) \htext{$\circ$}
\textref h:C v:C
\move (976 796) \htext{$\circ$}
\textref h:C v:C
\move (985 797) \htext{$\circ$}
\textref h:C v:C
\move (1004 798) \htext{$\circ$}
\textref h:C v:C
\move (1013 795) \htext{$\circ$}
\textref h:C v:C
\move (1022 793) \htext{$\circ$}
\textref h:C v:C
\move (1031 798) \htext{$\circ$}
\textref h:C v:C
\move (1040 798) \htext{$\circ$}
\textref h:C v:C
\move (1049 795) \htext{$\circ$}
\textref h:C v:C
\move (1057 793) \htext{$\circ$}
\textref h:C v:C
\move (1074 799) \htext{$\circ$}
\textref h:C v:C
\move (1083 794) \htext{$\circ$}
\textref h:C v:C
\move (1091 794) \htext{$\circ$}
\textref h:C v:C
\move (1099 797) \htext{$\circ$}
\textref h:C v:C
\move (1107 797) \htext{$\circ$}
\textref h:C v:C
\move (1116 796) \htext{$\circ$}
\textref h:C v:C
\move (1123 795) \htext{$\circ$}
\textref h:C v:C
\move (1139 797) \htext{$\circ$}
\textref h:C v:C
\move (1147 796) \htext{$\circ$}
\textref h:C v:C
\move (1162 790) \htext{$\circ$}
\textref h:C v:C
\move (1177 794) \htext{$\circ$}
\textref h:C v:C
\move (1185 795) \htext{$\circ$}
\textref h:C v:C
\move (1207 799) \htext{$\circ$}
\textref h:C v:C
\move (1214 797) \htext{$\circ$}
\textref h:C v:C
\move (1228 795) \htext{$\circ$}
\textref h:C v:C
\move (1242 794) \htext{$\circ$}
\textref h:C v:C
\move (1256 791) \htext{$\circ$}
\textref h:C v:C
\move (1263 794) \htext{$\circ$}
\textref h:C v:C
\move (1277 792) \htext{$\circ$}
\textref h:C v:C
\move (1283 803) \htext{$\circ$}
\textref h:C v:C
\move (1290 794) \htext{$\circ$}
\textref h:C v:C
\move (1316 794) \htext{$\circ$}
\textref h:C v:C
\move (1348 794) \htext{$\circ$}
\textref h:C v:C
\move (1361 791) \htext{$\circ$}
\textref h:C v:C
\move (1415 800) \htext{$\circ$}
\textref h:C v:C
\move (1479 792) \htext{$\circ$}
\textref h:C v:C
\move(650 320) \htext{$\circ$}
\textref h:L v:C
\move(695 320) \htext{$t=5$
}
\textref h:C v:C
\textref h:C v:C
\move (266 227) \htext{$\Box$}
\textref h:C v:C
\move (383 660) \htext{$\Box$}
\textref h:C v:C
\move (431 730) \htext{$\Box$}
\textref h:C v:C
\move (468 737) \htext{$\Box$}
\textref h:C v:C
\move (500 761) \htext{$\Box$}
\textref h:C v:C
\move (527 768) \htext{$\Box$}
\textref h:C v:C
\move (552 780) \htext{$\Box$}
\textref h:C v:C
\move (596 783) \htext{$\Box$}
\textref h:C v:C
\move (616 792) \htext{$\Box$}
\textref h:C v:C
\move (635 794) \htext{$\Box$}
\textref h:C v:C
\move (653 790) \htext{$\Box$}
\textref h:C v:C
\move (670 791) \htext{$\Box$}
\textref h:C v:C
\move (687 781) \htext{$\Box$}
\textref h:C v:C
\move (703 783) \htext{$\Box$}
\textref h:C v:C
\move (733 815) \htext{$\Box$}
\textref h:C v:C
\move (747 795) \htext{$\Box$}
\textref h:C v:C
\move (761 794) \htext{$\Box$}
\textref h:C v:C
\move (775 803) \htext{$\Box$}
\textref h:C v:C
\move (788 785) \htext{$\Box$}
\textref h:C v:C
\move (801 799) \htext{$\Box$}
\textref h:C v:C
\move (813 811) \htext{$\Box$}
\textref h:C v:C
\move (838 804) \htext{$\Box$}
\textref h:C v:C
\move (850 794) \htext{$\Box$}
\textref h:C v:C
\move (861 793) \htext{$\Box$}
\textref h:C v:C
\move (872 792) \htext{$\Box$}
\textref h:C v:C
\move (894 794) \htext{$\Box$}
\textref h:C v:C
\move (905 790) \htext{$\Box$}
\textref h:C v:C
\move (926 802) \htext{$\Box$}
\textref h:C v:C
\move (936 800) \htext{$\Box$}
\textref h:C v:C
\move (946 798) \htext{$\Box$}
\textref h:C v:C
\move (956 798) \htext{$\Box$}
\textref h:C v:C
\move (966 794) \htext{$\Box$}
\textref h:C v:C
\move (976 778) \htext{$\Box$}
\textref h:C v:C
\move (985 796) \htext{$\Box$}
\textref h:C v:C
\move (1004 790) \htext{$\Box$}
\textref h:C v:C
\move (1013 794) \htext{$\Box$}
\textref h:C v:C
\move (1022 787) \htext{$\Box$}
\textref h:C v:C
\move (1031 811) \htext{$\Box$}
\textref h:C v:C
\move (1040 791) \htext{$\Box$}
\textref h:C v:C
\move (1049 788) \htext{$\Box$}
\textref h:C v:C
\move (1057 790) \htext{$\Box$}
\textref h:C v:C
\move (1074 791) \htext{$\Box$}
\textref h:C v:C
\move (1083 785) \htext{$\Box$}
\textref h:C v:C
\move (1091 788) \htext{$\Box$}
\textref h:C v:C
\move (1099 803) \htext{$\Box$}
\textref h:C v:C
\move (1107 794) \htext{$\Box$}
\textref h:C v:C
\move (1116 799) \htext{$\Box$}
\textref h:C v:C
\move (1123 796) \htext{$\Box$}
\textref h:C v:C
\move (1139 794) \htext{$\Box$}
\textref h:C v:C
\move (1147 792) \htext{$\Box$}
\textref h:C v:C
\move (1162 776) \htext{$\Box$}
\textref h:C v:C
\move (1177 792) \htext{$\Box$}
\textref h:C v:C
\move (1185 794) \htext{$\Box$}
\textref h:C v:C
\move (1207 801) \htext{$\Box$}
\textref h:C v:C
\move (1214 793) \htext{$\Box$}
\textref h:C v:C
\move (1228 793) \htext{$\Box$}
\textref h:C v:C
\move (1242 791) \htext{$\Box$}
\textref h:C v:C
\move (1256 799) \htext{$\Box$}
\textref h:C v:C
\move (1263 800) \htext{$\Box$}
\textref h:C v:C
\move (1277 789) \htext{$\Box$}
\textref h:C v:C
\move (1283 784) \htext{$\Box$}
\textref h:C v:C
\move (1290 794) \htext{$\Box$}
\textref h:C v:C
\move (1316 803) \htext{$\Box$}
\textref h:C v:C
\move (1348 804) \htext{$\Box$}
\textref h:C v:C
\move (1361 795) \htext{$\Box$}
\textref h:C v:C
\move (1415 817) \htext{$\Box$}
\textref h:C v:C
\move (1479 818) \htext{$\Box$}
\textref h:C v:C
\move(650 275) \htext{$\Box$}
\textref h:L v:C
\move(695 275) \htext{$t=6$
}
\textref h:C v:C
\end{texdraw}}
\end{center}

%% file: fig4.tex
\begin{center}                                  
                \makebox[0pt]{\begin{texdraw}
                \normalsize
                \ifx\pathDEFINED\relax\else\let\pathDEFINED\relax
                 \def\QtGfr{\ifx (\TGre \let\YhetT\cpath\else\let\YhetT\relax\fi\YhetT}
                 \def\path (#1 #2){\move (#1 #2)\futurelet\TGre\QtGfr}
                 \def\cpath (#1 #2){\lvec (#1 #2)\futurelet\TGre\QtGfr}
                \fi
                \drawdim pt
                \setunitscale 0.216 
                \linewd 3
\special{PSfile=fig4.1.ps hscale=90 vscale=90}
\move(850 25) \textref h:C v:C \htext{distance $r$}
\move(25 510) \textref h:C v:C \vtext{}
\linewd 3
\textref h:C v:C
\move (266 75) \htext{0}
\move (500 75) \htext{2}
\move (733 75) \htext{4}
\move (966 75) \htext{6}
\move (1200 75) \htext{8}
\move (1433 75) \htext{10}
\textref h:R v:C
\move (135 243) \htext{$-0.020$}
\move (135 376) \htext{$-0.015$}
\move (135 509) \htext{$-0.010$}
\move (135 643) \htext{$-0.005$}
\move (135 776) \htext{$0.000$}
\textref h:C v:C
\move(800 950) \textref h:C v:C \htext{$E^0(r,t), \kappa=0.14077$}
\textref h:C v:C
\move (266 343) \htext{$\bullet$}
\textref h:C v:C
\move (383 571) \htext{$\bullet$}
\textref h:C v:C
\move (431 665) \htext{$\bullet$}
\textref h:C v:C
\move (468 710) \htext{$\bullet$}
\textref h:C v:C
\move (500 716) \htext{$\bullet$}
\textref h:C v:C
\move (527 738) \htext{$\bullet$}
\textref h:C v:C
\move (552 750) \htext{$\bullet$}
\textref h:C v:C
\move (596 760) \htext{$\bullet$}
\textref h:C v:C
\move (616 764) \htext{$\bullet$}
\textref h:C v:C
\move (635 764) \htext{$\bullet$}
\textref h:C v:C
\move (653 768) \htext{$\bullet$}
\textref h:C v:C
\move (670 769) \htext{$\bullet$}
\textref h:C v:C
\move (687 769) \htext{$\bullet$}
\textref h:C v:C
\move (703 772) \htext{$\bullet$}
\textref h:C v:C
\move (733 769) \htext{$\bullet$}
\textref h:C v:C
\move (747 773) \htext{$\bullet$}
\textref h:C v:C
\move (761 773) \htext{$\bullet$}
\textref h:C v:C
\move (775 774) \htext{$\bullet$}
\textref h:C v:C
\move (788 774) \htext{$\bullet$}
\textref h:C v:C
\move (801 774) \htext{$\bullet$}
\textref h:C v:C
\move (813 774) \htext{$\bullet$}
\textref h:C v:C
\move (838 773) \htext{$\bullet$}
\textref h:C v:C
\move (850 774) \htext{$\bullet$}
\textref h:C v:C
\move (861 775) \htext{$\bullet$}
\textref h:C v:C
\move (872 775) \htext{$\bullet$}
\textref h:C v:C
\move (894 776) \htext{$\bullet$}
\textref h:C v:C
\move (905 775) \htext{$\bullet$}
\textref h:C v:C
\move (926 778) \htext{$\bullet$}
\textref h:C v:C
\move (936 776) \htext{$\bullet$}
\textref h:C v:C
\move (946 775) \htext{$\bullet$}
\textref h:C v:C
\move (956 776) \htext{$\bullet$}
\textref h:C v:C
\move (966 775) \htext{$\bullet$}
\textref h:C v:C
\move (976 774) \htext{$\bullet$}
\textref h:C v:C
\move (985 776) \htext{$\bullet$}
\textref h:C v:C
\move (1004 777) \htext{$\bullet$}
\textref h:C v:C
\move (1013 776) \htext{$\bullet$}
\textref h:C v:C
\move (1022 777) \htext{$\bullet$}
\textref h:C v:C
\move (1031 777) \htext{$\bullet$}
\textref h:C v:C
\move (1040 777) \htext{$\bullet$}
\textref h:C v:C
\move (1049 777) \htext{$\bullet$}
\textref h:C v:C
\move (1057 777) \htext{$\bullet$}
\textref h:C v:C
\move (1074 778) \htext{$\bullet$}
\textref h:C v:C
\move (1083 778) \htext{$\bullet$}
\textref h:C v:C
\move (1091 777) \htext{$\bullet$}
\textref h:C v:C
\move (1099 778) \htext{$\bullet$}
\textref h:C v:C
\move (1107 775) \htext{$\bullet$}
\textref h:C v:C
\move (1116 776) \htext{$\bullet$}
\textref h:C v:C
\move (1123 776) \htext{$\bullet$}
\textref h:C v:C
\move (1139 777) \htext{$\bullet$}
\textref h:C v:C
\move (1147 777) \htext{$\bullet$}
\textref h:C v:C
\move (1162 776) \htext{$\bullet$}
\textref h:C v:C
\move (1177 777) \htext{$\bullet$}
\textref h:C v:C
\move (1185 777) \htext{$\bullet$}
\textref h:C v:C
\move (1207 774) \htext{$\bullet$}
\textref h:C v:C
\move (1214 776) \htext{$\bullet$}
\textref h:C v:C
\move (1228 777) \htext{$\bullet$}
\textref h:C v:C
\move (1242 777) \htext{$\bullet$}
\textref h:C v:C
\move (1256 774) \htext{$\bullet$}
\textref h:C v:C
\move (1263 774) \htext{$\bullet$}
\textref h:C v:C
\move (1277 777) \htext{$\bullet$}
\textref h:C v:C
\move (1283 777) \htext{$\bullet$}
\textref h:C v:C
\move (1290 776) \htext{$\bullet$}
\textref h:C v:C
\move (1316 774) \htext{$\bullet$}
\textref h:C v:C
\move (1348 775) \htext{$\bullet$}
\textref h:C v:C
\move (1361 776) \htext{$\bullet$}
\textref h:C v:C
\move (1415 770) \htext{$\bullet$}
\textref h:C v:C
\move (1479 771) \htext{$\bullet$}
\textref h:C v:C
\move(650 410) \htext{$\bullet$}
\textref h:L v:C
\move(695 410) \htext{$t=3$
}
\textref h:C v:C
\textref h:C v:C
\move (266 423) \htext{$\Diamond$}
\textref h:C v:C
\move (383 611) \htext{$\Diamond$}
\textref h:C v:C
\move (431 687) \htext{$\Diamond$}
\textref h:C v:C
\move (468 719) \htext{$\Diamond$}
\textref h:C v:C
\move (500 722) \htext{$\Diamond$}
\textref h:C v:C
\move (527 743) \htext{$\Diamond$}
\textref h:C v:C
\move (552 750) \htext{$\Diamond$}
\textref h:C v:C
\move (596 760) \htext{$\Diamond$}
\textref h:C v:C
\move (616 765) \htext{$\Diamond$}
\textref h:C v:C
\move (635 768) \htext{$\Diamond$}
\textref h:C v:C
\move (653 772) \htext{$\Diamond$}
\textref h:C v:C
\move (670 772) \htext{$\Diamond$}
\textref h:C v:C
\move (687 767) \htext{$\Diamond$}
\textref h:C v:C
\move (703 770) \htext{$\Diamond$}
\textref h:C v:C
\move (733 773) \htext{$\Diamond$}
\textref h:C v:C
\move (747 773) \htext{$\Diamond$}
\textref h:C v:C
\move (761 778) \htext{$\Diamond$}
\textref h:C v:C
\move (775 775) \htext{$\Diamond$}
\textref h:C v:C
\move (788 779) \htext{$\Diamond$}
\textref h:C v:C
\move (801 774) \htext{$\Diamond$}
\textref h:C v:C
\move (813 770) \htext{$\Diamond$}
\textref h:C v:C
\move (838 774) \htext{$\Diamond$}
\textref h:C v:C
\move (850 772) \htext{$\Diamond$}
\textref h:C v:C
\move (861 775) \htext{$\Diamond$}
\textref h:C v:C
\move (872 776) \htext{$\Diamond$}
\textref h:C v:C
\move (894 773) \htext{$\Diamond$}
\textref h:C v:C
\move (905 774) \htext{$\Diamond$}
\textref h:C v:C
\move (926 774) \htext{$\Diamond$}
\textref h:C v:C
\move (936 776) \htext{$\Diamond$}
\textref h:C v:C
\move (946 776) \htext{$\Diamond$}
\textref h:C v:C
\move (956 776) \htext{$\Diamond$}
\textref h:C v:C
\move (966 772) \htext{$\Diamond$}
\textref h:C v:C
\move (976 771) \htext{$\Diamond$}
\textref h:C v:C
\move (985 775) \htext{$\Diamond$}
\textref h:C v:C
\move (1004 771) \htext{$\Diamond$}
\textref h:C v:C
\move (1013 775) \htext{$\Diamond$}
\textref h:C v:C
\move (1022 776) \htext{$\Diamond$}
\textref h:C v:C
\move (1031 779) \htext{$\Diamond$}
\textref h:C v:C
\move (1040 776) \htext{$\Diamond$}
\textref h:C v:C
\move (1049 779) \htext{$\Diamond$}
\textref h:C v:C
\move (1057 778) \htext{$\Diamond$}
\textref h:C v:C
\move (1074 778) \htext{$\Diamond$}
\textref h:C v:C
\move (1083 780) \htext{$\Diamond$}
\textref h:C v:C
\move (1091 780) \htext{$\Diamond$}
\textref h:C v:C
\move (1099 780) \htext{$\Diamond$}
\textref h:C v:C
\move (1107 777) \htext{$\Diamond$}
\textref h:C v:C
\move (1116 775) \htext{$\Diamond$}
\textref h:C v:C
\move (1123 778) \htext{$\Diamond$}
\textref h:C v:C
\move (1139 782) \htext{$\Diamond$}
\textref h:C v:C
\move (1147 775) \htext{$\Diamond$}
\textref h:C v:C
\move (1162 777) \htext{$\Diamond$}
\textref h:C v:C
\move (1177 776) \htext{$\Diamond$}
\textref h:C v:C
\move (1185 779) \htext{$\Diamond$}
\textref h:C v:C
\move (1207 775) \htext{$\Diamond$}
\textref h:C v:C
\move (1214 780) \htext{$\Diamond$}
\textref h:C v:C
\move (1228 778) \htext{$\Diamond$}
\textref h:C v:C
\move (1242 780) \htext{$\Diamond$}
\textref h:C v:C
\move (1256 778) \htext{$\Diamond$}
\textref h:C v:C
\move (1263 775) \htext{$\Diamond$}
\textref h:C v:C
\move (1277 781) \htext{$\Diamond$}
\textref h:C v:C
\move (1283 773) \htext{$\Diamond$}
\textref h:C v:C
\move (1290 778) \htext{$\Diamond$}
\textref h:C v:C
\move (1316 776) \htext{$\Diamond$}
\textref h:C v:C
\move (1348 778) \htext{$\Diamond$}
\textref h:C v:C
\move (1361 778) \htext{$\Diamond$}
\textref h:C v:C
\move (1415 774) \htext{$\Diamond$}
\textref h:C v:C
\move (1479 768) \htext{$\Diamond$}
\textref h:C v:C
\move(650 365) \htext{$\Diamond$}
\textref h:L v:C
\move(695 365) \htext{$t=4$
}
\textref h:C v:C
\textref h:C v:C
\move (266 508) \htext{$\circ$}
\textref h:C v:C
\move (383 642) \htext{$\circ$}
\textref h:C v:C
\move (431 696) \htext{$\circ$}
\textref h:C v:C
\move (468 696) \htext{$\circ$}
\textref h:C v:C
\move (500 689) \htext{$\circ$}
\textref h:C v:C
\move (527 732) \htext{$\circ$}
\textref h:C v:C
\move (552 729) \htext{$\circ$}
\textref h:C v:C
\move (596 727) \htext{$\circ$}
\textref h:C v:C
\move (616 747) \htext{$\circ$}
\textref h:C v:C
\move (635 766) \htext{$\circ$}
\textref h:C v:C
\move (653 767) \htext{$\circ$}
\textref h:C v:C
\move (670 746) \htext{$\circ$}
\textref h:C v:C
\move (687 759) \htext{$\circ$}
\textref h:C v:C
\move (703 765) \htext{$\circ$}
\textref h:C v:C
\move (733 763) \htext{$\circ$}
\textref h:C v:C
\move (747 770) \htext{$\circ$}
\textref h:C v:C
\move (761 769) \htext{$\circ$}
\textref h:C v:C
\move (775 775) \htext{$\circ$}
\textref h:C v:C
\move (788 778) \htext{$\circ$}
\textref h:C v:C
\move (801 768) \htext{$\circ$}
\textref h:C v:C
\move (813 778) \htext{$\circ$}
\textref h:C v:C
\move (838 773) \htext{$\circ$}
\textref h:C v:C
\move (850 762) \htext{$\circ$}
\textref h:C v:C
\move (861 771) \htext{$\circ$}
\textref h:C v:C
\move (872 782) \htext{$\circ$}
\textref h:C v:C
\move (894 770) \htext{$\circ$}
\textref h:C v:C
\move (905 776) \htext{$\circ$}
\textref h:C v:C
\move (926 761) \htext{$\circ$}
\textref h:C v:C
\move (936 759) \htext{$\circ$}
\textref h:C v:C
\move (946 767) \htext{$\circ$}
\textref h:C v:C
\move (956 772) \htext{$\circ$}
\textref h:C v:C
\move (966 762) \htext{$\circ$}
\textref h:C v:C
\move (976 770) \htext{$\circ$}
\textref h:C v:C
\move (985 774) \htext{$\circ$}
\textref h:C v:C
\move (1004 762) \htext{$\circ$}
\textref h:C v:C
\move (1013 764) \htext{$\circ$}
\textref h:C v:C
\move (1022 771) \htext{$\circ$}
\textref h:C v:C
\move (1031 793) \htext{$\circ$}
\textref h:C v:C
\move (1040 786) \htext{$\circ$}
\textref h:C v:C
\move (1049 778) \htext{$\circ$}
\textref h:C v:C
\move (1057 769) \htext{$\circ$}
\textref h:C v:C
\move (1074 771) \htext{$\circ$}
\textref h:C v:C
\move (1083 773) \htext{$\circ$}
\textref h:C v:C
\move (1091 780) \htext{$\circ$}
\textref h:C v:C
\move (1099 777) \htext{$\circ$}
\textref h:C v:C
\move (1107 756) \htext{$\circ$}
\textref h:C v:C
\move (1116 776) \htext{$\circ$}
\textref h:C v:C
\move (1123 774) \htext{$\circ$}
\textref h:C v:C
\move (1139 787) \htext{$\circ$}
\textref h:C v:C
\move (1147 769) \htext{$\circ$}
\textref h:C v:C
\move (1162 779) \htext{$\circ$}
\textref h:C v:C
\move (1177 773) \htext{$\circ$}
\textref h:C v:C
\move (1185 782) \htext{$\circ$}
\textref h:C v:C
\move (1207 777) \htext{$\circ$}
\textref h:C v:C
\move (1214 780) \htext{$\circ$}
\textref h:C v:C
\move (1228 784) \htext{$\circ$}
\textref h:C v:C
\move (1242 775) \htext{$\circ$}
\textref h:C v:C
\move (1256 786) \htext{$\circ$}
\textref h:C v:C
\move (1263 775) \htext{$\circ$}
\textref h:C v:C
\move (1277 768) \htext{$\circ$}
\textref h:C v:C
\move (1283 773) \htext{$\circ$}
\textref h:C v:C
\move (1290 786) \htext{$\circ$}
\textref h:C v:C
\move (1316 756) \htext{$\circ$}
\textref h:C v:C
\move (1348 774) \htext{$\circ$}
\textref h:C v:C
\move (1361 805) \htext{$\circ$}
\textref h:C v:C
\move (1415 805) \htext{$\circ$}
\textref h:C v:C
\move (1479 847) \htext{$\circ$}
\textref h:C v:C
\move(650 320) \htext{$\circ$}
\textref h:L v:C
\move(695 320) \htext{$t=5$
}
\textref h:C v:C
\textref h:C v:C
\move (266 260) \htext{$\Box$}
\textref h:C v:C
\move (383 649) \htext{$\Box$}
\textref h:C v:C
\move (431 739) \htext{$\Box$}
\textref h:C v:C
\move (468 662) \htext{$\Box$}
\textref h:C v:C
\move (500 798) \htext{$\Box$}
\textref h:C v:C
\move (527 761) \htext{$\Box$}
\textref h:C v:C
\move (552 724) \htext{$\Box$}
\textref h:C v:C
\move (596 694) \htext{$\Box$}
\textref h:C v:C
\move (616 755) \htext{$\Box$}
\textref h:C v:C
\move (635 782) \htext{$\Box$}
\textref h:C v:C
\move (653 798) \htext{$\Box$}
\textref h:C v:C
\move (670 724) \htext{$\Box$}
\textref h:C v:C
\move (687 809) \htext{$\Box$}
\textref h:C v:C
\move (703 803) \htext{$\Box$}
\textref h:C v:C
\move (733 817) \htext{$\Box$}
\textref h:C v:C
\move (747 805) \htext{$\Box$}
\textref h:C v:C
\move (761 773) \htext{$\Box$}
\textref h:C v:C
\move (775 741) \htext{$\Box$}
\textref h:C v:C
\move (788 849) \htext{$\Box$}
\textref h:C v:C
\move (801 777) \htext{$\Box$}
\textref h:C v:C
\move (813 798) \htext{$\Box$}
\textref h:C v:C
\move (838 777) \htext{$\Box$}
\textref h:C v:C
\move (850 776) \htext{$\Box$}
\textref h:C v:C
\move (861 764) \htext{$\Box$}
\textref h:C v:C
\move (872 820) \htext{$\Box$}
\textref h:C v:C
\move (894 731) \htext{$\Box$}
\textref h:C v:C
\move (905 756) \htext{$\Box$}
\textref h:C v:C
\move (926 724) \htext{$\Box$}
\textref h:C v:C
\move (936 694) \htext{$\Box$}
\textref h:C v:C
\move (946 748) \htext{$\Box$}
\textref h:C v:C
\move (956 741) \htext{$\Box$}
\textref h:C v:C
\move (966 738) \htext{$\Box$}
\textref h:C v:C
\move (976 797) \htext{$\Box$}
\textref h:C v:C
\move (985 739) \htext{$\Box$}
\textref h:C v:C
\move (1004 774) \htext{$\Box$}
\textref h:C v:C
\move (1013 763) \htext{$\Box$}
\textref h:C v:C
\move (1022 730) \htext{$\Box$}
\textref h:C v:C
\move (1031 715) \htext{$\Box$}
\textref h:C v:C
\move (1040 702) \htext{$\Box$}
\textref h:C v:C
\move (1049 794) \htext{$\Box$}
\textref h:C v:C
\move (1057 761) \htext{$\Box$}
\textref h:C v:C
\move (1074 861) \htext{$\Box$}
\textref h:C v:C
\move (1083 707) \htext{$\Box$}
\textref h:C v:C
\move (1091 772) \htext{$\Box$}
\textref h:C v:C
\move (1099 874) \htext{$\Box$}
\textref h:C v:C
\move (1107 698) \htext{$\Box$}
\textref h:C v:C
\move (1116 760) \htext{$\Box$}
\textref h:C v:C
\move (1123 766) \htext{$\Box$}
\textref h:C v:C
\move (1139 812) \htext{$\Box$}
\textref h:C v:C
\move (1147 751) \htext{$\Box$}
\textref h:C v:C
\move (1162 735) \htext{$\Box$}
\textref h:C v:C
\move (1177 763) \htext{$\Box$}
\textref h:C v:C
\move (1185 831) \htext{$\Box$}
\textref h:C v:C
\move (1207 824) \htext{$\Box$}
\textref h:C v:C
\move (1214 708) \htext{$\Box$}
\textref h:C v:C
\move (1228 788) \htext{$\Box$}
\textref h:C v:C
\move (1242 776) \htext{$\Box$}
\textref h:C v:C
\move (1256 738) \htext{$\Box$}
\textref h:C v:C
\move (1263 780) \htext{$\Box$}
\textref h:C v:C
\move (1277 672) \htext{$\Box$}
\textref h:C v:C
\move (1283 707) \htext{$\Box$}
\textref h:C v:C
\move (1290 772) \htext{$\Box$}
\textref h:C v:C
\move (1316 759) \htext{$\Box$}
\textref h:C v:C
\move (1348 821) \htext{$\Box$}
\textref h:C v:C
\move (1361 838) \htext{$\Box$}
\textref h:C v:C
\move (1415 866) \htext{$\Box$}
\textref h:C v:C
\textref h:C v:C
\move(650 275) \htext{$\Box$}
\textref h:L v:C
\move(695 275) \htext{$t=6$
}
\textref h:C v:C
\end{texdraw}}
\end{center}
\vspace{1em}
\begin{center}                                  
                \makebox[0pt]{\begin{texdraw}
                \normalsize
                \ifx\pathDEFINED\relax\else\let\pathDEFINED\relax
                 \def\QtGfr{\ifx (\TGre \let\YhetT\cpath\else\let\YhetT\relax\fi\YhetT}
                 \def\path (#1 #2){\move (#1 #2)\futurelet\TGre\QtGfr}
                 \def\cpath (#1 #2){\lvec (#1 #2)\futurelet\TGre\QtGfr}
                \fi
                \drawdim pt
                \setunitscale 0.216 
                \linewd 3
\special{PSfile=fig4.2.ps hscale=90 vscale=90}
\move(850 25) \textref h:C v:C \htext{distance $r$}
\move(25 510) \textref h:C v:C \vtext{}
\linewd 3
\textref h:C v:C
\move (266 75) \htext{0}
\move (500 75) \htext{2}
\move (733 75) \htext{4}
\move (966 75) \htext{6}
\move (1200 75) \htext{8}
\move (1433 75) \htext{10}
\textref h:R v:C
\move (135 224) \htext{$-0.10$}
\move (135 338) \htext{$-0.08$}
\move (135 452) \htext{$-0.06$}
\move (135 567) \htext{$-0.04$}
\move (135 681) \htext{$-0.02$}
\move (135 795) \htext{$0.00$}
\textref h:C v:C
\move(800 950) \textref h:C v:C \htext{$\overline{E}(r,t), \kappa=0.14077$}
\textref h:C v:C
\move (266 329) \htext{$\bullet$}
\textref h:C v:C
\move (383 690) \htext{$\bullet$}
\textref h:C v:C
\move (431 762) \htext{$\bullet$}
\textref h:C v:C
\move (468 782) \htext{$\bullet$}
\textref h:C v:C
\move (500 774) \htext{$\bullet$}
\textref h:C v:C
\move (527 786) \htext{$\bullet$}
\textref h:C v:C
\move (552 790) \htext{$\bullet$}
\textref h:C v:C
\move (596 792) \htext{$\bullet$}
\textref h:C v:C
\move (616 792) \htext{$\bullet$}
\textref h:C v:C
\move (635 791) \htext{$\bullet$}
\textref h:C v:C
\move (653 793) \htext{$\bullet$}
\textref h:C v:C
\move (670 794) \htext{$\bullet$}
\textref h:C v:C
\move (687 794) \htext{$\bullet$}
\textref h:C v:C
\move (703 794) \htext{$\bullet$}
\textref h:C v:C
\move (733 793) \htext{$\bullet$}
\textref h:C v:C
\move (747 794) \htext{$\bullet$}
\textref h:C v:C
\move (761 794) \htext{$\bullet$}
\textref h:C v:C
\move (775 794) \htext{$\bullet$}
\textref h:C v:C
\move (788 794) \htext{$\bullet$}
\textref h:C v:C
\move (801 795) \htext{$\bullet$}
\textref h:C v:C
\move (813 795) \htext{$\bullet$}
\textref h:C v:C
\move (838 795) \htext{$\bullet$}
\textref h:C v:C
\move (850 795) \htext{$\bullet$}
\textref h:C v:C
\move (861 795) \htext{$\bullet$}
\textref h:C v:C
\move (872 795) \htext{$\bullet$}
\textref h:C v:C
\move (894 795) \htext{$\bullet$}
\textref h:C v:C
\move (905 795) \htext{$\bullet$}
\textref h:C v:C
\move (926 795) \htext{$\bullet$}
\textref h:C v:C
\move (936 795) \htext{$\bullet$}
\textref h:C v:C
\move (946 795) \htext{$\bullet$}
\textref h:C v:C
\move (956 795) \htext{$\bullet$}
\textref h:C v:C
\move (966 795) \htext{$\bullet$}
\textref h:C v:C
\move (976 795) \htext{$\bullet$}
\textref h:C v:C
\move (985 795) \htext{$\bullet$}
\textref h:C v:C
\move (1004 795) \htext{$\bullet$}
\textref h:C v:C
\move (1013 795) \htext{$\bullet$}
\textref h:C v:C
\move (1022 795) \htext{$\bullet$}
\textref h:C v:C
\move (1031 795) \htext{$\bullet$}
\textref h:C v:C
\move (1040 795) \htext{$\bullet$}
\textref h:C v:C
\move (1049 795) \htext{$\bullet$}
\textref h:C v:C
\move (1057 795) \htext{$\bullet$}
\textref h:C v:C
\move (1074 796) \htext{$\bullet$}
\textref h:C v:C
\move (1083 795) \htext{$\bullet$}
\textref h:C v:C
\move (1091 795) \htext{$\bullet$}
\textref h:C v:C
\move (1099 795) \htext{$\bullet$}
\textref h:C v:C
\move (1107 795) \htext{$\bullet$}
\textref h:C v:C
\move (1116 795) \htext{$\bullet$}
\textref h:C v:C
\move (1123 795) \htext{$\bullet$}
\textref h:C v:C
\move (1139 795) \htext{$\bullet$}
\textref h:C v:C
\move (1147 795) \htext{$\bullet$}
\textref h:C v:C
\move (1162 795) \htext{$\bullet$}
\textref h:C v:C
\move (1177 795) \htext{$\bullet$}
\textref h:C v:C
\move (1185 795) \htext{$\bullet$}
\textref h:C v:C
\move (1207 795) \htext{$\bullet$}
\textref h:C v:C
\move (1214 795) \htext{$\bullet$}
\textref h:C v:C
\move (1228 795) \htext{$\bullet$}
\textref h:C v:C
\move (1242 795) \htext{$\bullet$}
\textref h:C v:C
\move (1256 795) \htext{$\bullet$}
\textref h:C v:C
\move (1263 796) \htext{$\bullet$}
\textref h:C v:C
\move (1277 794) \htext{$\bullet$}
\textref h:C v:C
\move (1283 795) \htext{$\bullet$}
\textref h:C v:C
\move (1290 795) \htext{$\bullet$}
\textref h:C v:C
\move (1316 795) \htext{$\bullet$}
\textref h:C v:C
\move (1348 795) \htext{$\bullet$}
\textref h:C v:C
\move (1361 795) \htext{$\bullet$}
\textref h:C v:C
\move (1415 795) \htext{$\bullet$}
\textref h:C v:C
\move (1479 795) \htext{$\bullet$}
\textref h:C v:C
\move(650 410) \htext{$\bullet$}
\textref h:L v:C
\move(695 410) \htext{$t=3$
}
\textref h:C v:C
\textref h:C v:C
\move (266 306) \htext{$\Diamond$}
\textref h:C v:C
\move (383 694) \htext{$\Diamond$}
\textref h:C v:C
\move (431 758) \htext{$\Diamond$}
\textref h:C v:C
\move (468 775) \htext{$\Diamond$}
\textref h:C v:C
\move (500 770) \htext{$\Diamond$}
\textref h:C v:C
\move (527 783) \htext{$\Diamond$}
\textref h:C v:C
\move (552 788) \htext{$\Diamond$}
\textref h:C v:C
\move (596 790) \htext{$\Diamond$}
\textref h:C v:C
\move (616 791) \htext{$\Diamond$}
\textref h:C v:C
\move (635 789) \htext{$\Diamond$}
\textref h:C v:C
\move (653 793) \htext{$\Diamond$}
\textref h:C v:C
\move (670 791) \htext{$\Diamond$}
\textref h:C v:C
\move (687 792) \htext{$\Diamond$}
\textref h:C v:C
\move (703 794) \htext{$\Diamond$}
\textref h:C v:C
\move (733 793) \htext{$\Diamond$}
\textref h:C v:C
\move (747 794) \htext{$\Diamond$}
\textref h:C v:C
\move (761 794) \htext{$\Diamond$}
\textref h:C v:C
\move (775 794) \htext{$\Diamond$}
\textref h:C v:C
\move (788 794) \htext{$\Diamond$}
\textref h:C v:C
\move (801 795) \htext{$\Diamond$}
\textref h:C v:C
\move (813 794) \htext{$\Diamond$}
\textref h:C v:C
\move (838 794) \htext{$\Diamond$}
\textref h:C v:C
\move (850 795) \htext{$\Diamond$}
\textref h:C v:C
\move (861 795) \htext{$\Diamond$}
\textref h:C v:C
\move (872 796) \htext{$\Diamond$}
\textref h:C v:C
\move (894 795) \htext{$\Diamond$}
\textref h:C v:C
\move (905 796) \htext{$\Diamond$}
\textref h:C v:C
\move (926 796) \htext{$\Diamond$}
\textref h:C v:C
\move (936 795) \htext{$\Diamond$}
\textref h:C v:C
\move (946 795) \htext{$\Diamond$}
\textref h:C v:C
\move (956 795) \htext{$\Diamond$}
\textref h:C v:C
\move (966 795) \htext{$\Diamond$}
\textref h:C v:C
\move (976 794) \htext{$\Diamond$}
\textref h:C v:C
\move (985 795) \htext{$\Diamond$}
\textref h:C v:C
\move (1004 794) \htext{$\Diamond$}
\textref h:C v:C
\move (1013 795) \htext{$\Diamond$}
\textref h:C v:C
\move (1022 795) \htext{$\Diamond$}
\textref h:C v:C
\move (1031 794) \htext{$\Diamond$}
\textref h:C v:C
\move (1040 796) \htext{$\Diamond$}
\textref h:C v:C
\move (1049 795) \htext{$\Diamond$}
\textref h:C v:C
\move (1057 796) \htext{$\Diamond$}
\textref h:C v:C
\move (1074 795) \htext{$\Diamond$}
\textref h:C v:C
\move (1083 795) \htext{$\Diamond$}
\textref h:C v:C
\move (1091 796) \htext{$\Diamond$}
\textref h:C v:C
\move (1099 796) \htext{$\Diamond$}
\textref h:C v:C
\move (1107 797) \htext{$\Diamond$}
\textref h:C v:C
\move (1116 794) \htext{$\Diamond$}
\textref h:C v:C
\move (1123 794) \htext{$\Diamond$}
\textref h:C v:C
\move (1139 794) \htext{$\Diamond$}
\textref h:C v:C
\move (1147 796) \htext{$\Diamond$}
\textref h:C v:C
\move (1162 794) \htext{$\Diamond$}
\textref h:C v:C
\move (1177 795) \htext{$\Diamond$}
\textref h:C v:C
\move (1185 796) \htext{$\Diamond$}
\textref h:C v:C
\move (1207 795) \htext{$\Diamond$}
\textref h:C v:C
\move (1214 795) \htext{$\Diamond$}
\textref h:C v:C
\move (1228 794) \htext{$\Diamond$}
\textref h:C v:C
\move (1242 795) \htext{$\Diamond$}
\textref h:C v:C
\move (1256 797) \htext{$\Diamond$}
\textref h:C v:C
\move (1263 796) \htext{$\Diamond$}
\textref h:C v:C
\move (1277 795) \htext{$\Diamond$}
\textref h:C v:C
\move (1283 795) \htext{$\Diamond$}
\textref h:C v:C
\move (1290 795) \htext{$\Diamond$}
\textref h:C v:C
\move (1316 794) \htext{$\Diamond$}
\textref h:C v:C
\move (1348 795) \htext{$\Diamond$}
\textref h:C v:C
\move (1361 793) \htext{$\Diamond$}
\textref h:C v:C
\move (1415 794) \htext{$\Diamond$}
\textref h:C v:C
\move (1479 796) \htext{$\Diamond$}
\textref h:C v:C
\move(650 365) \htext{$\Diamond$}
\textref h:L v:C
\move(695 365) \htext{$t=4$
}
\textref h:C v:C
\textref h:C v:C
\move (266 314) \htext{$\circ$}
\textref h:C v:C
\move (383 695) \htext{$\circ$}
\textref h:C v:C
\move (431 751) \htext{$\circ$}
\textref h:C v:C
\move (468 766) \htext{$\circ$}
\textref h:C v:C
\move (500 768) \htext{$\circ$}
\textref h:C v:C
\move (527 779) \htext{$\circ$}
\textref h:C v:C
\move (552 785) \htext{$\circ$}
\textref h:C v:C
\move (596 791) \htext{$\circ$}
\textref h:C v:C
\move (616 789) \htext{$\circ$}
\textref h:C v:C
\move (635 790) \htext{$\circ$}
\textref h:C v:C
\move (653 798) \htext{$\circ$}
\textref h:C v:C
\move (670 786) \htext{$\circ$}
\textref h:C v:C
\move (687 794) \htext{$\circ$}
\textref h:C v:C
\move (703 795) \htext{$\circ$}
\textref h:C v:C
\move (733 795) \htext{$\circ$}
\textref h:C v:C
\move (747 791) \htext{$\circ$}
\textref h:C v:C
\move (761 795) \htext{$\circ$}
\textref h:C v:C
\move (775 795) \htext{$\circ$}
\textref h:C v:C
\move (788 792) \htext{$\circ$}
\textref h:C v:C
\move (801 795) \htext{$\circ$}
\textref h:C v:C
\move (813 793) \htext{$\circ$}
\textref h:C v:C
\move (838 794) \htext{$\circ$}
\textref h:C v:C
\move (850 792) \htext{$\circ$}
\textref h:C v:C
\move (861 796) \htext{$\circ$}
\textref h:C v:C
\move (872 797) \htext{$\circ$}
\textref h:C v:C
\move (894 794) \htext{$\circ$}
\textref h:C v:C
\move (905 798) \htext{$\circ$}
\textref h:C v:C
\move (926 797) \htext{$\circ$}
\textref h:C v:C
\move (936 796) \htext{$\circ$}
\textref h:C v:C
\move (946 793) \htext{$\circ$}
\textref h:C v:C
\move (956 798) \htext{$\circ$}
\textref h:C v:C
\move (966 794) \htext{$\circ$}
\textref h:C v:C
\move (976 795) \htext{$\circ$}
\textref h:C v:C
\move (985 797) \htext{$\circ$}
\textref h:C v:C
\move (1004 795) \htext{$\circ$}
\textref h:C v:C
\move (1013 795) \htext{$\circ$}
\textref h:C v:C
\move (1022 796) \htext{$\circ$}
\textref h:C v:C
\move (1031 793) \htext{$\circ$}
\textref h:C v:C
\move (1040 792) \htext{$\circ$}
\textref h:C v:C
\move (1049 795) \htext{$\circ$}
\textref h:C v:C
\move (1057 796) \htext{$\circ$}
\textref h:C v:C
\move (1074 794) \htext{$\circ$}
\textref h:C v:C
\move (1083 797) \htext{$\circ$}
\textref h:C v:C
\move (1091 794) \htext{$\circ$}
\textref h:C v:C
\move (1099 795) \htext{$\circ$}
\textref h:C v:C
\move (1107 800) \htext{$\circ$}
\textref h:C v:C
\move (1116 796) \htext{$\circ$}
\textref h:C v:C
\move (1123 795) \htext{$\circ$}
\textref h:C v:C
\move (1139 793) \htext{$\circ$}
\textref h:C v:C
\move (1147 793) \htext{$\circ$}
\textref h:C v:C
\move (1162 794) \htext{$\circ$}
\textref h:C v:C
\move (1177 793) \htext{$\circ$}
\textref h:C v:C
\move (1185 797) \htext{$\circ$}
\textref h:C v:C
\move (1207 795) \htext{$\circ$}
\textref h:C v:C
\move (1214 795) \htext{$\circ$}
\textref h:C v:C
\move (1228 790) \htext{$\circ$}
\textref h:C v:C
\move (1242 796) \htext{$\circ$}
\textref h:C v:C
\move (1256 799) \htext{$\circ$}
\textref h:C v:C
\move (1263 801) \htext{$\circ$}
\textref h:C v:C
\move (1277 798) \htext{$\circ$}
\textref h:C v:C
\move (1283 799) \htext{$\circ$}
\textref h:C v:C
\move (1290 792) \htext{$\circ$}
\textref h:C v:C
\move (1316 800) \htext{$\circ$}
\textref h:C v:C
\move (1348 795) \htext{$\circ$}
\textref h:C v:C
\move (1361 798) \htext{$\circ$}
\textref h:C v:C
\move (1415 794) \htext{$\circ$}
\textref h:C v:C
\move (1479 789) \htext{$\circ$}
\textref h:C v:C
\move(650 320) \htext{$\circ$}
\textref h:L v:C
\move(695 320) \htext{$t=5$
}
\textref h:C v:C
\textref h:C v:C
\move (266 369) \htext{$\Box$}
\textref h:C v:C
\move (383 703) \htext{$\Box$}
\textref h:C v:C
\move (431 737) \htext{$\Box$}
\textref h:C v:C
\move (468 739) \htext{$\Box$}
\textref h:C v:C
\move (500 782) \htext{$\Box$}
\textref h:C v:C
\move (527 776) \htext{$\Box$}
\textref h:C v:C
\move (552 774) \htext{$\Box$}
\textref h:C v:C
\move (596 792) \htext{$\Box$}
\textref h:C v:C
\move (616 779) \htext{$\Box$}
\textref h:C v:C
\move (635 776) \htext{$\Box$}
\textref h:C v:C
\move (653 787) \htext{$\Box$}
\textref h:C v:C
\move (670 765) \htext{$\Box$}
\textref h:C v:C
\move (687 786) \htext{$\Box$}
\textref h:C v:C
\move (703 789) \htext{$\Box$}
\textref h:C v:C
\move (733 802) \htext{$\Box$}
\textref h:C v:C
\move (747 794) \htext{$\Box$}
\textref h:C v:C
\move (761 788) \htext{$\Box$}
\textref h:C v:C
\move (775 791) \htext{$\Box$}
\textref h:C v:C
\move (788 784) \htext{$\Box$}
\textref h:C v:C
\move (801 791) \htext{$\Box$}
\textref h:C v:C
\move (813 798) \htext{$\Box$}
\textref h:C v:C
\move (838 795) \htext{$\Box$}
\textref h:C v:C
\move (850 788) \htext{$\Box$}
\textref h:C v:C
\move (861 791) \htext{$\Box$}
\textref h:C v:C
\move (872 783) \htext{$\Box$}
\textref h:C v:C
\move (894 793) \htext{$\Box$}
\textref h:C v:C
\move (905 784) \htext{$\Box$}
\textref h:C v:C
\move (926 790) \htext{$\Box$}
\textref h:C v:C
\move (936 791) \htext{$\Box$}
\textref h:C v:C
\move (946 786) \htext{$\Box$}
\textref h:C v:C
\move (956 801) \htext{$\Box$}
\textref h:C v:C
\move (966 794) \htext{$\Box$}
\textref h:C v:C
\move (976 789) \htext{$\Box$}
\textref h:C v:C
\move (985 797) \htext{$\Box$}
\textref h:C v:C
\move (1004 769) \htext{$\Box$}
\textref h:C v:C
\move (1013 790) \htext{$\Box$}
\textref h:C v:C
\move (1022 800) \htext{$\Box$}
\textref h:C v:C
\move (1031 788) \htext{$\Box$}
\textref h:C v:C
\move (1040 796) \htext{$\Box$}
\textref h:C v:C
\move (1049 800) \htext{$\Box$}
\textref h:C v:C
\move (1057 800) \htext{$\Box$}
\textref h:C v:C
\move (1074 816) \htext{$\Box$}
\textref h:C v:C
\move (1083 786) \htext{$\Box$}
\textref h:C v:C
\move (1091 787) \htext{$\Box$}
\textref h:C v:C
\move (1099 779) \htext{$\Box$}
\textref h:C v:C
\move (1107 814) \htext{$\Box$}
\textref h:C v:C
\move (1116 794) \htext{$\Box$}
\textref h:C v:C
\move (1123 786) \htext{$\Box$}
\textref h:C v:C
\move (1139 799) \htext{$\Box$}
\textref h:C v:C
\move (1147 803) \htext{$\Box$}
\textref h:C v:C
\move (1162 777) \htext{$\Box$}
\textref h:C v:C
\move (1177 788) \htext{$\Box$}
\textref h:C v:C
\move (1185 792) \htext{$\Box$}
\textref h:C v:C
\move (1207 786) \htext{$\Box$}
\textref h:C v:C
\move (1214 787) \htext{$\Box$}
\textref h:C v:C
\move (1228 787) \htext{$\Box$}
\textref h:C v:C
\move (1242 793) \htext{$\Box$}
\textref h:C v:C
\move (1256 787) \htext{$\Box$}
\textref h:C v:C
\move (1263 803) \htext{$\Box$}
\textref h:C v:C
\move (1277 800) \htext{$\Box$}
\textref h:C v:C
\move (1283 800) \htext{$\Box$}
\textref h:C v:C
\move (1290 782) \htext{$\Box$}
\textref h:C v:C
\move (1316 809) \htext{$\Box$}
\textref h:C v:C
\move (1348 784) \htext{$\Box$}
\textref h:C v:C
\move (1361 789) \htext{$\Box$}
\textref h:C v:C
\move (1415 774) \htext{$\Box$}
\textref h:C v:C
\move (1479 765) \htext{$\Box$}
\textref h:C v:C
\move(650 275) \htext{$\Box$}
\textref h:L v:C
\move(695 275) \htext{$t=6$
}
\textref h:C v:C
\end{texdraw}}
\end{center}

%% file: fig5.tex
\begin{center}                                  
                \makebox[0pt]{\begin{texdraw}
                \normalsize
                \ifx\pathDEFINED\relax\else\let\pathDEFINED\relax
                 \def\QtGfr{\ifx (\TGre \let\YhetT\cpath\else\let\YhetT\relax\fi\YhetT}
                 \def\path (#1 #2){\move (#1 #2)\futurelet\TGre\QtGfr}
                 \def\cpath (#1 #2){\lvec (#1 #2)\futurelet\TGre\QtGfr}
                \fi
                \drawdim pt
                \setunitscale 0.216 
                \linewd 3
\special{PSfile=fig5.1.ps hscale=90 vscale=90}
\move(850 25) \textref h:C v:C \htext{$r$}
\move(25 510) \textref h:C v:C \vtext{$\log \overline E$}
\linewd 3
\textref h:C v:C
\move (227 75) \htext{0.0}
\move (383 75) \htext{0.5}
\move (538 75) \htext{1.0}
\move (694 75) \htext{1.5}
\move (850 75) \htext{2.0}
\move (1005 75) \htext{2.5}
\move (1161 75) \htext{3.0}
\move (1316 75) \htext{3.5}
\move (1472 75) \htext{4.0}
\textref h:R v:C
\move (135 160) \htext{$-9$}
\move (135 260) \htext{$-8$}
\move (135 360) \htext{$-7$}
\move (135 460) \htext{$-6$}
\move (135 560) \htext{$-5$}
\move (135 660) \htext{$-4$}
\move (135 760) \htext{$-3$}
\move (135 860) \htext{$-2$}
\textref h:C v:C
\move(800 950) \textref h:C v:C \htext{}
\textref h:C v:C
\move (227 823) \htext{$\bullet$}
\textref h:C v:C
\move (538 668) \htext{$\bullet$}
\textref h:C v:C
\move (667 572) \htext{$\bullet$}
\textref h:C v:C
\move (766 503) \htext{$\bullet$}
\textref h:C v:C
\move (850 519) \htext{$\bullet$}
\textref h:C v:C
\move (923 461) \htext{$\bullet$}
\textref h:C v:C
\move (989 415) \htext{$\bullet$}
\textref h:C v:C
\move (1107 376) \htext{$\bullet$}
\textref h:C v:C
\move (1161 359) \htext{$\bullet$}
\textref h:C v:C
\move (1211 300) \htext{$\bullet$}
\textref h:C v:C
\move (1259 280) \htext{$\bullet$}
\textref h:C v:C
\move (1305 256) \htext{$\bullet$}
\textref h:C v:C
\move (1349 247) \htext{$\bullet$}
\textref h:C v:C
\move(210 210) \htext{$\bullet$}
\textref h:L v:C
\move(255 210) \htext{$\kappa_1$, $g^{\mathrm{L}}=0.535(0.045)$, $\chi^2=2.365$}
\textref h:C v:C
\end{texdraw}}
\end{center}

%% file: fig6.tex
\begin{center}                                  
                \makebox[0pt]{\begin{texdraw}
                \normalsize
                \ifx\pathDEFINED\relax\else\let\pathDEFINED\relax
                 \def\QtGfr{\ifx (\TGre \let\YhetT\cpath\else\let\YhetT\relax\fi\YhetT}
                 \def\path (#1 #2){\move (#1 #2)\futurelet\TGre\QtGfr}
                 \def\cpath (#1 #2){\lvec (#1 #2)\futurelet\TGre\QtGfr}
                \fi
                \drawdim pt
                \setunitscale 0.216 
                \linewd 3
\special{PSfile=fig6.1.ps hscale=90 vscale=90}
\move(850 25) \textref h:C v:C \htext{$r$}
\move(25 510) \textref h:C v:C \vtext{$\log \overline E$}
\linewd 3
\textref h:C v:C
\move (227 75) \htext{0.0}
\move (383 75) \htext{0.5}
\move (538 75) \htext{1.0}
\move (694 75) \htext{1.5}
\move (850 75) \htext{2.0}
\move (1005 75) \htext{2.5}
\move (1161 75) \htext{3.0}
\move (1316 75) \htext{3.5}
\move (1472 75) \htext{4.0}
\textref h:R v:C
\move (135 160) \htext{$-9$}
\move (135 260) \htext{$-8$}
\move (135 360) \htext{$-7$}
\move (135 460) \htext{$-6$}
\move (135 560) \htext{$-5$}
\move (135 660) \htext{$-4$}
\move (135 760) \htext{$-3$}
\move (135 860) \htext{$-2$}
\textref h:C v:C
\move(800 950) \textref h:C v:C \htext{}
\textref h:C v:C
\move (227 813) \htext{$\bullet$}
\textref h:C v:C
\move (538 656) \htext{$\bullet$}
\textref h:C v:C
\move (667 561) \htext{$\bullet$}
\textref h:C v:C
\move (766 502) \htext{$\bullet$}
\textref h:C v:C
\move (850 519) \htext{$\bullet$}
\textref h:C v:C
\move (923 450) \htext{$\bullet$}
\textref h:C v:C
\move (989 397) \htext{$\bullet$}
\textref h:C v:C
\move (1107 352) \htext{$\bullet$}
\textref h:C v:C
\move (1161 351) \htext{$\bullet$}
\textref h:C v:C
\move (1211 371) \htext{$\bullet$}
\textref h:C v:C
\move (1259 249) \htext{$\bullet$}
\textref h:C v:C
\move (1305 346) \htext{$\bullet$}
\textref h:C v:C
\move (1349 288) \htext{$\bullet$}
\textref h:C v:C
\move(210 210) \htext{$\bullet$}
\textref h:L v:C
\move(255 210) \htext{$\kappa_2$, $g^{\mathrm{L}}=0.527(0.047)$, $\chi^2=4.139$}
\textref h:C v:C
\end{texdraw}}
\end{center}